\documentclass[a4paper,11pt]{article}
\pdfoutput=1 

\usepackage{jcappub, amsmath, amssymb, framed} 
\newcommand{\bfv}{{\boldsymbol{v}}}
\newcommand{\bfk}{{\boldsymbol{k}}}

\newcommand{\bfpsi}{{\boldsymbol{\psi}}}

\newcommand{\bfx}{\boldsymbol{x}}

\newcommand{\nhat}{{\hat{\boldsymbol{n}}}}

\newcommand{\Msun}{{M_{\odot}}}

\def\refereeedit{}

\author[a]{Matthew McQuinn}
\author[b]{Anson D'Aloisio}

\affiliation[a]{Department of Astronomy, University of Washington,
Seattle, WA 98195}
\affiliation[b]{Department of Physics \& Astronomy, University of California, Riverside, CA 92521, USA}

\emailAdd{mcquinn@uw.edu}

\title{the observable 21cm signal from reionization may be perturbative}

\abstract{      
We develop an effective perturbation theory (and, equivalently, a bias expansion) for the inhomogeneous 21cm radiation field from reionization.  Using large-scale simulations of cosmological reionization, we find that this expansion describes the modes in the simulated 21cm signal over much of the wavenumber range probed by upcoming 21cm arrays.  This result provides an understanding of the potential signal shapes that compliments the nonlinear numerical modeling that has been the focus of most previous work.  We find that the observable signal often can be described with $2-3$ bias coefficients that can be interpreted in terms of the source biases, the average neutral fraction, the characteristic size of ionized regions, and the patchiness of reionization.  The 21cm signal serves as a wacky example of a bias expansion in cosmology, with our approach synthesizing key results.
}


\begin{document}
\maketitle
\section{introduction}

Reionization is an astrophysically complex era in which the first sources of ionizing emissions turned on and ionized all of the neutral hydrogen aside from the small fraction that lies within galaxies (see \cite{mcquinn-review} for a recent review).   Analyses of the Ly$\alpha$ forest show that reionization is largely complete by $z\gtrsim 6$ \citep{mcgreer15}, and cosmic microwave background measurements yield a mean redshift of $z\approx 7-8$ \citep{2016arXiv160503507P}.  Better measurements of the reionization process would further constrain the properties of the ionizing sources \citep{furl-models, mcquinn07, 2017MNRAS.469.4283K}, the abundance of small-scale structures that absorb many of the ionizing photons \citep{ciardi06, 2009MNRAS.394..960C, furlanetto07b}, and perhaps even cosmological parameters \citep{barkana04, mcquinn06, 2008PhRvD..78b3529M, 2013MNRAS.433.2900D, 2013PhRvD..88h1303M, 2013PhRvD..88b3534L}.

By directly imaging the neutral hydrogen as it was reionized, high-redshift 21cm observations have more potential than other observations to constrain the structure of reionization   \citep{madau97, ciardi03-21cm, zaldarriaga04, furl-rev}. In the past decade, there has been a worldwide effort to refine redshifted 21cm instrumentation and analysis methods \citep[e.g.][]{morales03, bowman05, mcquinn06, 2011PhRvD..83j3006L, parsons12, dillon15, 2016ApJ...819....8P, 2016MNRAS.461.3135B, 2017MNRAS.470.1849E}.  While at present these efforts have yielded only upper bounds on the signal, these bounds have been improving steadily \citep{2016ApJ...833..102B,  2017ApJ...838...65P}. In the coming years, the Hydrogen Epoch of Reionization Array \citep[HERA;][]{2017PASP..129d5001D} and eventually the Square Kilometer Array \citep{2015aska.confE...1K} are anticipated to provide the first detection of this signal.  Once a detection is claimed, the challenges will be to (1) confirm whether the signal is indeed redshifted 21cm radiation and not residual foreground contamination and to (2) interpret what the signal means for reionization.  

Our most reliable method for understanding the reionization process is computationally expensive radiative transfer simulations \citep{gnedin97, ciardi03-sim, iliev05, mcquinn07, trac07, 2014ApJ...793...30G, 2017MNRAS.466..960P}, but only a handful of simulations have been run in the $>100~$Mpc box sizes that are required to capture a representative sample of structures \citep{barkana04, iliev05, mcquinn07,trac07}.  These simulations have difficulty capturing the smallest intergalactic structures that may act as sinks of ionizing photons and retard the reionization process \citep{miralda00, gnedin00, shapiro03}, and they are too computationally expensive to survey the vast parameter space of potential source models.  Computationally inexpensive semi-analytic reionization models, such as {\it 21CMFAST} \citep{mesinger11} and {\it SimFast21} \citep{2010MNRAS.406.2421S}, are able to meet the surveying challenge.  The semi-analytic models take a realization of the cosmological matter density distribution and place ionized regions around the locations where sources are likely to reside (and sometimes they also incorporate a prescription for the photon sinks; \cite{furlanetto05, alvarez12, sobacchi14}).  These semi-analytic algorithms have been tested against a few numerical simulations of reionization, showing agreement at the 50\% level \citep{zahn06, zahn11, 2014MNRAS.443.2843M}.  Here we develop a different method for understanding the 21cm signal.  This method may also be useful for understanding the semi-analytic models' signal shapes and the robustness of these models' predictions.

This paper shows that the Reionization Era 21cm signal is likely perturbative on many of the scales that existing and planned 21cm interferometric instruments are forecast to be sensitive.  On scales at which the signal is perturbative, it can be understood in terms of `bias' parameters that trace functions of the linear-theory matter density, with the values of these parameters often having straightforward interpretations.  The conclusion that this signal is perturbative is contrary to the pervading wisdom, which holds that the signal is highly nonlinear on the scales probed by 21cm telescopes.  
  That the signal is perturbative allows one to decompose the modes in the redshifted 21cm signal (and by extension any large-scale statistic applied to them) into functions of the linear-theory matter overdensity, $\delta^{(1)}$, ordered in powers of $\delta^{(1)}$.  

We are not the first to develop a perturbation theory for reionization; Zhang, Mao and collaborators \cite{2007MNRAS.375..324Z, 2015PhRvD..91h3015M} developed a linear perturbation theory of the form $T(k)\delta^{(1)}$.  To compute  $T(k)$, they assumed that nonlinear terms in their equations can be dropped without affecting the large-scale modes.  We expect that the large-scale morphology of reionization is quite sensitive to the neglected terms. For example, the small-scale clumpiness of neutral gas affects how far photons can travel and, hence, the sizes of the ionized bubbles.  Such an effect is described by the (previously dropped) nonlinear couplings between the ionization and the radiation.  We instead take an approach that consistently includes the large-scale effects of these terms.  We use that the `linear' transfer function must be analytic and so can be expanded as $T(k) = 1-  R_1^2 k^2 + R_2^2 k^4 +...$  (odd powers in $k$ violate statistical translational invariance), where the $R_i^2$ coefficients encapsulate the large-scale effects of small-scale nonlinear processes that perturbation theory cannot capture.  Just like the parameters in other effective theories,  
  understanding the meaning of this parameter requires an ultraviolet-complete model (in this case, semi-analytic models, reionization simulations, or the observations themselves).   In addition, we go beyond linear order, which is likely essential to explain the observable scales in the 21cm signal \citep{2007ApJ...659..865L}.  However, while going beyond linear order is essential, our key result is that much of the range of scales that upcoming observations could probe may be mildly nonlinear.  Our expansion is derived in two manners: (1) with a nonlocal cosmological biasing expansion and, almost equivalently, (2) starting with the equations of radiative transfer and developing an effective perturbation theory.  The free parameters of our theory encapsulate how reionization occurred, and, for some of them, we show that their approximate trends can be interpreted in terms of the source bias, the fraction of the Universe that is neutral, and the characteristic bubble size.


This paper is organized as follows.  Section~\ref{sec:evidence} uses the simulations of \cite{mcquinnLya} to argue that the 21 signal is likely perturbative over much of the wavenumber range that is accessible with upcoming observations.  We then write down a general bias expansion in Section~\ref{sec:bias}, and we develop an effective perturbation theory (EPT) expansion in Section~\ref{sec:PT}.  While equivalent to the bias expansion, the EPT approach motivates why some terms are likely small.  We show that this expansion (and even a much simplified version) successfully fits three very different reionization simulations in Section~\ref{sec:fit}; and we discuss physical interpretations for the fitting parameters.  Throughout, we use the standard Fourier convention in cosmology in which $2\pi$'s appear only under $dk$'s, and we will assume that all of reionization occurred during matter domination so that the growth factor scales with the scale factor, $a$.

While we were in the late stages of preparing this manuscript, \cite{2018arXiv180202578H} was submitted, showing that a quadratic bias model provides a good description of a configuration-space 21cm field generated by {\it 21CMFAST} when smoothed on $>30~$Mpc scales.  Our Fourier-space study reaches a similar conclusion, and, beyond this agreement, our study is quite distinct.

\section{evidence that the 21cm signal may be perturbative}
\label{sec:evidence}
\begin{figure}
\begin{center}
\epsfig{file=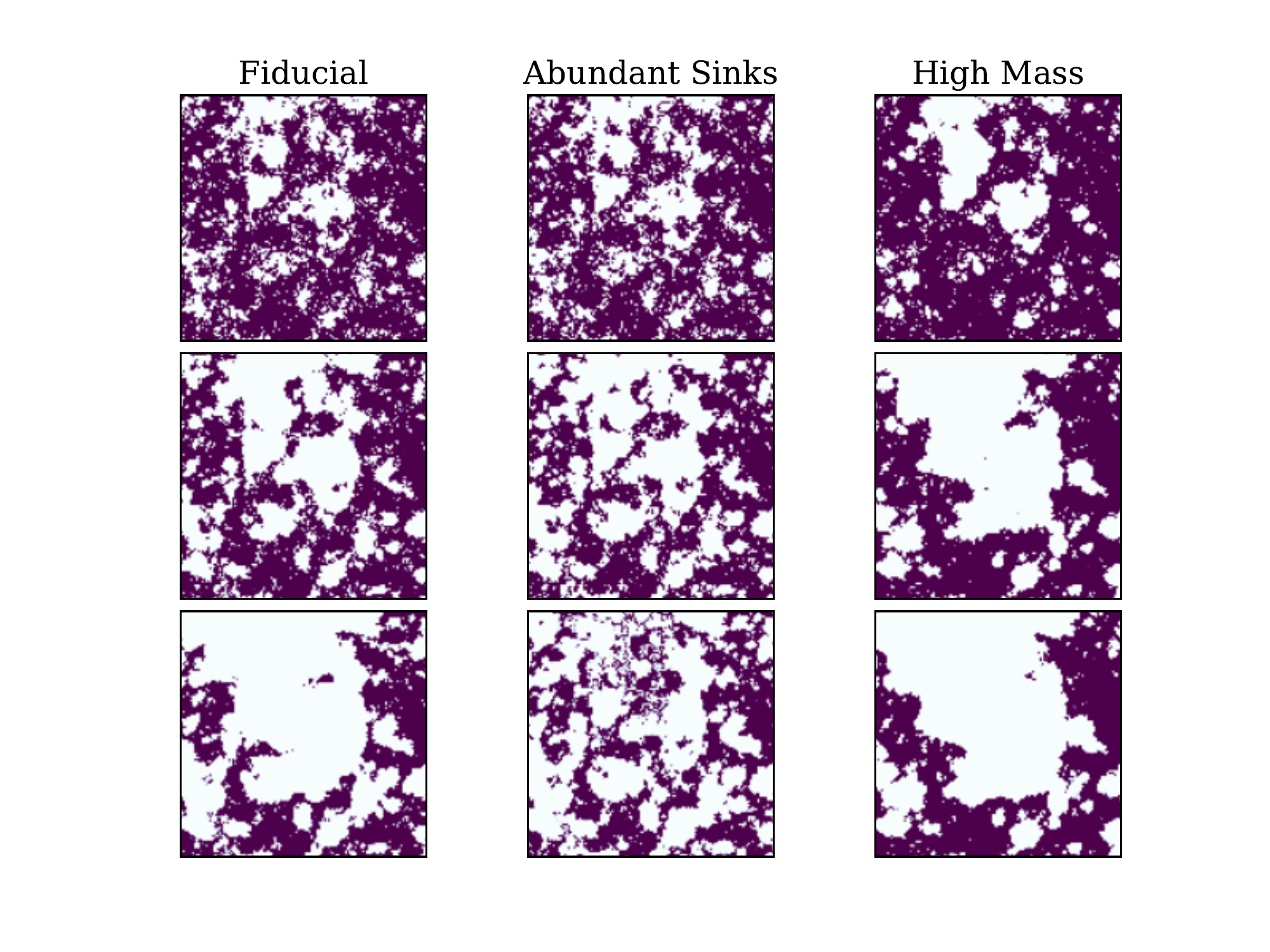, width=15.cm}
\vspace{-40pt}
\end{center}
\caption{Slices through the ionization field in our three $130\,h^{-1}\;$Mpc reionization simulations, showing from top to bottom snapshots with $\bar x_{\rm H} \approx 0.8$, $0.5$, $0.3$, originally run in \cite{mcquinn07, lidz08}.  White regions are ionized and black are neutral, and all simulations have the same same initial density field.  Even though the ionized structures can span many tens of comoving megaparsecs, over much of the wavenumber range that 21cm efforts are forecast to be sensitive we argue that the signal can be described perturbatively.  \label{fig:xill}}
\end{figure}    

This study uses three $130~h^{-1}$Mpc radiative transfer simulations of cosmological reionization, which are described in \cite{mcquinn07, lidz08}.  Sources of ionizing radiation are modeled in these simulations assuming a mapping between halo mass and luminosity (as these simulations do not capture galaxies).  The Fiducial simulation (the S1 simulation in \cite{mcquinn07}) assumes all halos more massive than $10^8\Msun$ produce ionizing photons with luminosity proportional to their halo mass, $L\propto M$.  
This model likely underestimates the clustering of reionization sources owing to the inefficiency of star formation in dwarf galaxies, with semi-analytic models of \cite{2017MNRAS.472.1576F} finding a super-linear mass-luminosity relation of $L\propto M^{1.2-1.5}$ above $10^8\Msun$, the exact index depending on the parametrization for stellar feedback. We also investigate a  simulation in which the sources are in higher mass halos with $L\propto M^{5/3}$ \cite[the S3 simulation in][]{mcquinn07}, a somewhat stronger scaling than found in the models of \cite{2017MNRAS.472.1576F}. Finally, our third `Abundant Sinks' simulation uses the same source model as the fiducial but models extra absorption (with the physical prescription based on unresolved minihalos \cite{shapiro03}, although any source of extra absorption should act to suppress the size of the largest bubbles; \cite{furlanetto05}).  We note that models for the Ly$\alpha$ forest immediately after reionization suggest that absorptions may play a prominent role in shaping reionization.\footnote{Models for the fluctuations in the ionizing background at $z=5-6$ require short mean free paths to explain the observed amplitude of opacity fluctuations in the Ly$\alpha$ forest \cite{becker15, 2016MNRAS.460.1328D, daloisio-gal}. In semi-analytic reionization calculations that use a single homogeneous mean free path chosen to match extrapolations from these models, the characteristic sizes of ionized bubbles are suppressed \cite{2014MNRAS.440.1662S}.
}   


Figure~\ref{fig:xill} shows slices through the ionization field in the three simulations, showing from top to bottom snapshots with $\bar x_{\rm H} \approx 0.8$, $0.5$, $0.3$.  White regions are ionized and black are neutral, and all simulations have the same same initial density field.   Most striking is the large ionized structures spanning many tens of megaparsecs.  These structures are the largest in the High Mass simulation, as the sources are most clustered in this case \citep{mcquinn07}, and smallest in the Abundant Sinks simulations, as absorptions preferentially act to reduce the growth of the largest ionized regions \citep{furlanetto05}. Towards the end of reionization, some of the ionized regions show sizes that extend a significant extent of the simulation box.  Because of these large ionized structures, the consensus view -- expressed by the almost sole concentration on numerical models -- is that the 21cm signal on observable scales is unsuitable for perturbative theory.

\begin{figure}
\begin{center}
\epsfig{file=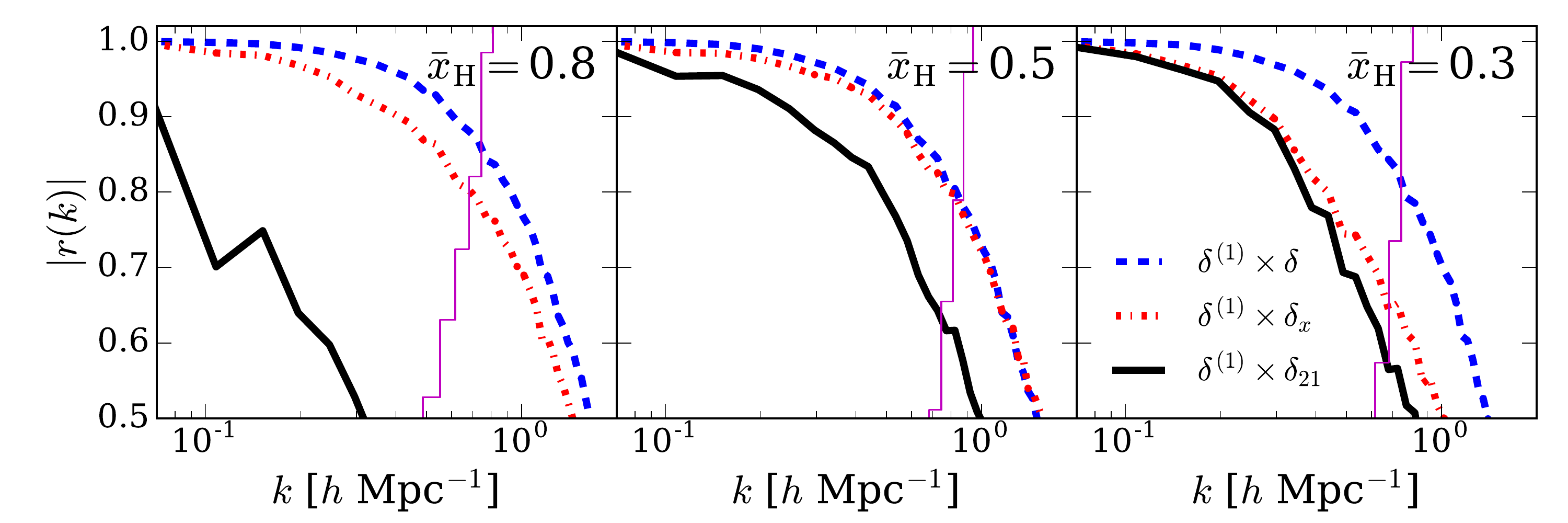, height=5.cm}
\end{center}
\vspace{-10pt}
\caption{The absolute value of the cross-correlation coefficient, $|r|$, between the linear matter overdensity $\delta^{(1)}$  and the overdensity in (1) the nonlinear matter $\delta$ (blue dashed curves), (2) the neutral hydrogen fraction $\delta_x$ (red dot-dashed curves), and (3) the 21cm signal $\delta_{21}$ (black solid curves), with all quantities computed from the fiducial simulation.  From left to right, the three panels correspond to global neutral fractions of $\bar x_{\rm H} = 0.8$, $0.5$ and $0.3$ and redshifts of $z=8.3$, $7.5$ and $7.0$.  The thin magenta segmented lines show the ratio of the projected HERA thermal noise at $z=8$ to the 21cm signal in bins of $\Delta k = 0.06\,h~$Mpc$^{-1}$; roughly, HERA is sensitive to wavenumbers that fall leftward of these lines.  That $|r| > 0.6-0.8$ leftward of these lines suggests that the observable signal is only mildly nonlinear. (The exception being the 21cm signal for $\bar x_{\rm H} = 0.8$, which we show later occurs because the linear bias is close to zero.)
 \label{fig:CCC}}
\end{figure}    
  
\begin{figure}
\begin{center}
\epsfig{file=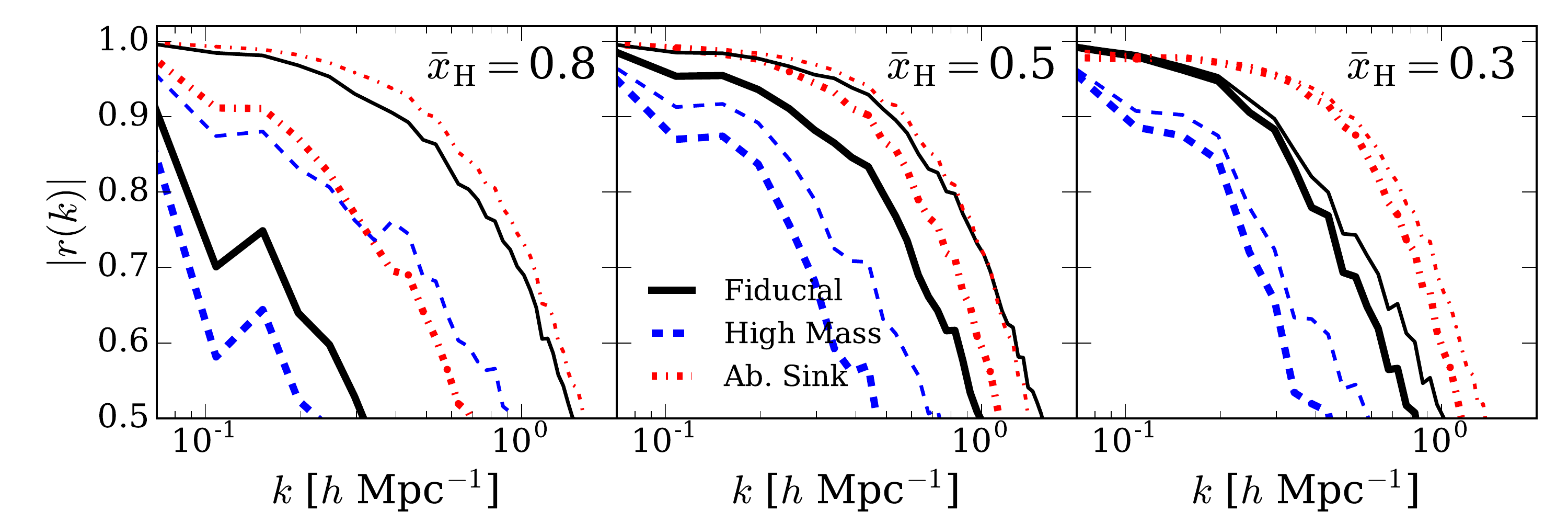, height=5.cm}
\end{center}
\vspace{-10pt}
\caption{The absolute value of the cross-correlation coefficient, $|r|$, between the linear matter overdensity  and (1) the neutral hydrogen fraction (thin curves) or (2) the 21cm signal assuming $T_S \gg T_{\rm CMB}$ (thick curves).  The solid curves show the Fiducial model, the dot-dashed curves show the Abundant Sinks model, and the dashed curves the High Mass model.  The magenta lines show the ratio of the projected HERA thermal noise at $z=8$ to the Fiducial simulation 21cm signal.
 \label{fig:CCCmodels}}
\end{figure}


Figure~\ref{fig:CCC} shows the absolute value of the cross-correlation coefficient, 
\begin{equation}
r(k) = \sum_{k-{\rm bin}} \frac{\delta_\bfk^{a}\delta_\bfk^{b*}}{\sqrt{\left |\delta_\bfk^{a} \right|^2 \left |\delta_\bfk^{b} \right|^2}},
\end{equation}
between different fields from our Fiducial reionization simulation. The different panels correspond to the snapshots with $(\bar x_{\rm H}, z)$ equal to $(0.8 ,8.3)$, $(0.5, 7.5)$ and $(0.3,7.0)$, from left to right respectively.   In this figure, $\delta_\bfk^{a}$ is the linear-theory (or initial) matter overdensity $\delta^{(1)}$, and $ \delta_\bfk^{b}$ is either (1) the nonlinear matter overdensity density $\delta$ [blue dashed curve], (2) the overdensity in the neutral hydrogen fraction $\delta_x$ [red dot-dashed curve], or (3) the 21cm `overdensity' given by $\delta_{21} \equiv x_{\rm H} (1+\delta)$ where $x_{\rm H}$ is the neutral hydrogen fraction [green dashed curve].  The 21cm signal traces $\delta_{21}$ to the extent that spin temperature fluctuations are unimportant, which traditionally has been found to be the case during the bulk of reionization \citep{furl-rev}, {\refereeedit although some more recent work has questioned this finding \citep{mirocha17}.  (We remark briefly on how spin temperature fluctuations might be incorporated into our formalism in the conclusions.)}   To the extent that $|r|= 1$, the phases of modes are the same as the initial phases in the matter density.  The thin magenta lines show the ratio of the $z=8$ HERA thermal noise to the 21cm power spectrum in bin sizes of $\Delta k = 0.06\,h~$Mpc$^{-1}$; HERA will be sensitive to modes leftward of these lines.  That $|r| > 0.6-0.8$ leftward of these lines suggests that the fields are only mildly nonlinear on scales that HERA probes.  (The exception being, surprisingly, the $\bar x_{\rm H} = 0.8$ snapshot.  We show later that when $\bar x_{\rm H} = 0.8$ the signal's linear bias is close to zero, which yields the small value for $|r|$.)  

Figure~\ref{fig:CCCmodels} again shows the absolute value of the cross-correlation coefficient, $|r|$, but for different reionization models.   The solid, dot-dashed, and dashed curves correspond to the Fiducial, Abundant Sinks, and High Mass simulations, respectively.    The thin curves are $|r|$ between the linear density, $\delta^{(1)}$, and the ionization, and the thick curves are between $\delta^{(1)}$ and the 21cm signal, $\delta_{21}$.  Generally $|r|$ is of similar magnitude between all the models:  It is somewhat larger in the Abundant Sinks simulation compared to the Fiducial simulation, whereas it is somewhat smaller in the High Mass simulation.  These trends owe to the sizes of the ionized bubbles, which are smallest in the Abundant Sinks simulation and largest in the High Mass one.

\section{the large-scale 21cm signal as a biased tracer of the matter}
\label{sec:bias}


  On large enough scales, everything -- including the 21cm signal from reionization -- is a biased tracer of local gravitational observables (the matter overdensity, the tidal field, the velocity divergence, etc; e.g. \cite{2016arXiv161109787D}).
  Previous work has determined the most general expansion possible \citep{2009JCAP...08..020M, 2014JCAP...08..056A, 2015JCAP...11..007S, 2016arXiv161109787D}:  One should include all scalar terms constructed from $\nabla_i \nabla_j \phi(x(t); t)$, where $\phi$ is the Newtonian gravitational potential, and its derivatives, as well as (the other independent scalar in a potential flow) the velocity potential $\phi_V \equiv \nabla \cdot \bfv$, evaluated at all prior times.  Here $x(t)$ tracks a fluid element's trajectory, as the properties that shape embedded tracers should be local functions of their path.  Let us keep all terms to third order except the one third-order term that depends on $\phi_V$, and we include the most relevant term that is higher order in derivatives.  With these choices (which will be justified) and up to shot noise, on large enough scales the signal must follow
 \begin{eqnarray}
\delta_{21}(t) &=& \int_0^{t} dt' b_1(t')\; \delta +   \int_0^{t} dt' b_{\nabla^2}(t')\; \nabla^2 \delta \\
&+& \int_0^{t} dt' b_2(t')\; \delta^2 + \overbrace{\int_0^{t} dt' b_{G2}(t')\; {\cal G}_2(\phi, \phi)}^{\rm likely~small;~we~keep}\\
&+& \overbrace{\int dt' b_3(t')\; \delta^3 +  \int_0^{t} dt' b_{G2\delta}(t')\; {\cal G}_2(\phi, \phi) \delta+\int_0^{t} dt' b_{G3}(t')\; {\cal G}_3}^{\rm not~needed~for~1-loop~power},
\label{eqn:expansion}
\end{eqnarray} 
where we have kept the $x(t)$ argument implicit and have adopted the basis set chosen in \citep{2014JCAP...08..056A} appropriate for Gaussian $\delta^{(1)}$.  Here  the `Galilean' operator ${\cal G}_2(\phi_1, \phi_2) \equiv \nabla_i \nabla_j \phi_1 \nabla^i \nabla^j \phi_2 - \nabla^2 \phi_1  \nabla^2 \phi_2$ \citep{2009JCAP...08..020M}, and ${\cal G}_3$ is a related higher-order Galilean operator \citep{2014JCAP...08..056A}.  Equation~(\ref{eqn:expansion}) includes all terms needed to calculate any statistic to 1-loop order.  {\refereeedit (For the purposes of this paper, which focuses on the power spectrum, it is sufficient to know that 1-loop order yields the first nonlinear correction to the tracer's power spectrum and each additional loop is needed for the next order correction.)}  In the last line, we have indicated the two terms that appear at third order in $\delta^{(1)}$ that we will subsequently drop because our focus is on the 1-loop power spectrum, where these terms do not contribute.  

We can reorganize our reduced 1-loop expansion in powers of in the linear matter overdensity, $\delta^{(1)}$, noting that $\delta^{(n)}$ is the $n^{\rm th}$-order density in Eulerian perturbation theory -- depending on $n$ powers of $\delta^{(1)}$.  Ordering in this way allows us to perform the time integrations in equation~(\ref{eqn:expansion}) using that $\delta^{(n)} \propto a^n$ at the redshifts of interest, yielding (up to a clarification given in the next paragraph) 
 \begin{eqnarray}
\delta_{21} &=&b_{1(1)} \delta^{(1)} + b_{\nabla^2(1)} \nabla^2  \delta^{(1)} \label{eqn:general bias}\\
 &+&b_{1(2)} \delta^{(2)} + b_{2(2)} [\delta^{(1)}]^2 +b_{G2(2)} {\cal G}_2(\phi^{(1)}, \phi^{(1)}) \nonumber \\
&+&b_{1(3)} \delta^{(3)}+2 b_{2(3)} [\delta^{(1)} \delta^{(2)}]+2 b_{G2(3)} {\cal G}_2(\phi^{(1)}, \phi^{(2)}) , \nonumber 
\end{eqnarray} 
where we have dropped the terms that we indicated are unimportant in equation~(\ref{eqn:expansion}) as well as the subcomponents of terms that do not contribute to the 1-loop power spectrum such as $b_{2(4)} [\delta^{(1)} \delta^{(3)}]$.  Additionally, we have kept only the linear order in $\nabla^2  \delta$ as motivated shortly. 
Here the notation is such that $b_{1(n)} \propto \int dt b_{1}(t) a(t)^n$.   

Two technical notes are in order.  First, all of the $\delta$ in equation~(\ref{eqn:expansion}) should be evaluated along fluid trajectories $\bfx(t')$ for $t'\leq t$, where $t$ is the present time.\footnote{Following fluid trajectories, required by locality, avoids infrared divergences that appear in ensemble-averaged quantities if, e.g., $\delta^{(2)}$ and $\delta^{(3)}$ enter with different biases.}  However, this would yield a more complex time dependence such that the integrations would not result in equation~(\ref{eqn:general bias}).   To yield something like equation~(\ref{eqn:general bias}), terms with different time dependences must be separated by expanding $\delta(\bfx(t')) = \delta(\bfx) - \partial_i \delta(\bfx) \Delta \psi_i(t') + ...$, where $\Delta\bfpsi \equiv \bfpsi(t) -  \bfpsi(t')$ is the displacement from time $t'$ to the position at the time of observation $t$  (evaluated to the desired order, with $\psi_i^{(n)}(t) \propto a(t)^n$), and similarly for other quantities, and adopting a notation where $\delta$ or $\Delta \Psi$ are evaluated at time $t$ if not given a temporal argument.  (At 1-loop, we only need to go to first order in $\Delta \psi_i$ since locally cubic terms can be absorbed into lower order terms.)   Collecting terms that are the same power in $a(t')$, which then yield the same bias coefficients, results in the substitutions $\delta^{(1)} \rightarrow \delta^{(1)} -  \partial_i \delta^{(1)} \psi_i$, $\delta^{(2)} \rightarrow \delta^{(2)} + \partial_i \delta^{(1)} \psi_i^{(1)} -  \partial_i \delta^{(2)} \psi_i$, etc. for equation~(\ref{eqn:general bias}) to be correct, computing the displacement $\bfpsi$ to the desired order. 
  These terms have the same time dependences as other terms and so no additional bias parameters are needed.  As a second technical note, $\delta^{(1)} \delta^{(2)}$, for example, depends on its small-scale smoothing, with different smoothings generating different levels of a large-scale term that scales as $\delta^{(1)}$ \citep{mcdonald06,2014JCAP...08..056A}.  We remove this UV sensitivity, `renormalizing' such quantities, following \cite{2014JCAP...08..056A}.  
  

%
%

 The above bias expansion keeps all terms that are relevant for the 1-loop power spectrum.  A second justification for the expansion comes from comparing the wavenumber scalings of all possible terms.  For a scale-free spectrum of density perturbations, each term's auto power spectrum (or its cross with other terms) scales as $(k/k_{\rm NL})^\gamma$ at $k \ll k_{\rm NL}$, where $k_{\rm NL}$ is the nonlinear wavenumber, with different terms having different $\gamma$.  Terms with $\gamma >0$ have stronger wavenumber scalings than shot noise, which motivates dropping them as these terms will likely be subdominant to shot noise at low wavenumbers (and, indeed, shot noise acts as a source of noise in our method for constraining various terms, making even the detection of terms with $\gamma >0$ challenging).  In the concordance cosmology, $n=\{-1.4, -2.2\}$ for $k=\{0.1, 0.5\}\,h$~Mpc$^{-1}$, the range of wavenumbers we consider.  Thus, our expansion should keep the term that scales with the linear power spectrum $P_L\sim k^n$ and perhaps the one that scales as $k^2 P_L \sim k^{n+2}$, but not terms with higher powers of real-space derivatives.  Indeed, keeping  $k^2 P_L$ is borderline, but since reionization is characterized by nonlocal effects, we keep it anyway.  Furthermore, since the $m$-loop order matter power spectrum in Eulerian perturbation theory scales like $P_{m-{\rm loop}} \propto k^{(m+1)(n+3)-3}$, keeping the $m=1$ order is justified as $P_{1-{\rm loop}} \propto k^{-1}$ for $n=-2$.  However, $P_{2-{\rm loop}} \propto k^{0}$ for $n=-2$ -- going beyond 1-loop order in the matter density is not motivated.  Considering the bias terms that are quadratic or higher powers in the matter density, $P_{[\delta^{(1)}]^2 \times [\delta^{(1)}]^2}$ scales as $k^{3+2n}$, which yields $k^{-1}$ for $n=-2$, as do similar terms involving the linear tidal fields, and so should be retained, but terms with higher powers of the matter density (that contribute to the power spectrum) are unlikely to be important. 
   

\section{an effective perturbation theory of reionization}
\label{sec:PT}

Here we develop an effective perturbation theory (EPT) for reionization, solving for the ionization field perturbatively using the radiative transfer equation plus the equation of ionization balance. {\refereeedit (An EPT solves perturbatively for a theory's low-wavenumber behavior, encapsulating the effects of small-scale non-perturbative processes by including all terms consistent with symmetry up to normalization constants that the EPT itself cannot predict.) }  Our EPT shows how such a biasing expansion arises from physical equations, motivates why certain terms in this expansion are likely large, and (with additional assumptions) allows us to develop simple expressions for the biases in terms of physical parameters (such as the neutral fraction and source biases).  The reader not interested in these intuition-developing results can safely skip to the next section, where we apply the bias expansion to the reionization simulations.

Let $I_\nu(\bfx, t, \nhat)$ be the proper ionizing photon specific intensity at comoving position $\bfx$, frequency $\nu$, direction $\nhat$, and time $t$.  The equation of radiative transfer is
\begin{equation}
\frac{\partial I_\nu}{\partial t} + \frac{\dot a}{a} \left[2 I_\nu - \nu \frac{\partial I_\nu}{\partial \nu}  \right] + a^{-1} \,c \,\nhat \cdot \nabla I_\nu = -c \kappa_\nu I_\nu + c j_\nu,
\label{eqn:RTeqn}
\end{equation}
where $j_\nu(\bfx, t)$ and  $\kappa_\nu(\bfx, t)$  are the specific emission and absorption coefficients (taken to be isotropic).
Because the speed of light is generally much greater than the velocities of ionization fronts \citep{daloisio18} and the sources evolve over a much longer timescale than the time it takes an emitted photon to be absorbed, we can treat the radiation field as time independent, and because the mean free path is much shorter than the Hubble scale ($c \kappa_\nu \gg H$), we can ignore redshifting. In these limits, the first two terms on the left-hand side of eqn.~(\ref{eqn:RTeqn}) can be dropped.   We also take the monochromatic limit, which we denote by dropping the frequency subscripts, because ionizing photons should be absorbed at sharp ionization fronts regardless of their frequency.  (This approximation would not apply if X-rays played an unexpectedly prominent role during reionization.) These approximations are also made by most radiative transfer simulations of reionization.  However, our final expressions, which re-derive the biasing expansion of the previous section, likely do not require them.  With these approximations and going to Fourier space, denoted with tildes, the radiative transfer equations becomes 
\begin{equation}
- i a^{-1} \nhat \cdot \bfk ~ \widetilde{I} = -  \widetilde{\kappa} \star \widetilde{I} +   \widetilde{j},
\label{eqn:RTeqnFourier}
\end{equation}
where the star denotes a convolution.

We would like to solve this equation (combined with an equation for ionization balance) perturbatively to calculate the ionization and 21cm fields.  However, we must be careful as, for example, the mean $\kappa$ to perturb around is unclear, as $\kappa$ can be essentially infinite in neutral regions and zero in ionized regions (and in regions where $\kappa$ is large $I$ is likely small).  We should instead expand around the mean of $\kappa I$, the absorption rate per unit volume.  The fields that affect the amount of absorption, $\kappa I$, are the neutral hydrogen fraction, the matter density, and the radiation intensity. Noting that the absorption is local in these fields, we may write
\begin{equation}
\kappa I =  \langle \kappa I \rangle \left(1 + b^{\kappa I}_{I, 1,} \delta_I + b^{\kappa I}_{x, 1} \delta_{x}  + b^{\kappa I}_{\delta, 1} \delta +  \sum_{X,Y \in \delta, \delta_{x},\delta_I} b^{\kappa I}_{XY, 2} [X Y - \langle XY \rangle] \right),
\label{eqn:kIexp}
\end{equation}
 where the $b^{\kappa I}_{X}$ are bias coefficients that we do \emph{not} attempt to calculate (as they are shaped by the very UV Jeans-scale clumping of the IGM), $\langle X \rangle$ denotes the volume average of $X$, $\delta_X$ is the overdensity in $X$.  {\it  Note that all overdensity fields in this section are (implicitly) smoothed on some scale where $\delta_X \ll 1$ so that the expansion in powers of $\delta_X$ converges.}   The above expression goes to second order as required for the $1$-loop power spectrum.  We ignore shot terms, defined as those uncorrelated with the large-scale fields.\footnote{Since the $\langle \kappa I \rangle$ (and later $\langle [ \Gamma x]  \rangle$) expansion is in the spatially smoothed fields rather than the unsmoothed fields, we should also include nonlocal terms of the form $k^2 \delta_X$.  These terms should be smaller than the nonlocal terms from the distance photons travel, $R$, and can be dropped (or, better yet, absorbed into $R$). \label{footn}}  


Writing eqn.~(\ref{eqn:kIexp}) in Fourier space and plugging it into eqn.~(\ref{eqn:RTeqnFourier}), we can solve for the overdensity in the intensity
\begin{equation}
b^{\kappa I}_{I,1} \widetilde{\delta_I}(\bfk, \nhat) =   \frac{ \widetilde{\delta_j}(\bfk) -b^{\kappa I}_{x, 1}\widetilde{\delta_{x}}(\bfk)  -b^{\kappa I}_{\delta, 1}\widetilde{\delta}(\bfk) - \sum_{X,Y \in \delta, \delta_{x},\delta_I} b^{\kappa I}_{XY, 2}~\widetilde{X} \star \widetilde{Y} }{1- i R \, \nhat \cdot \bfk},
\label{eqn:deltaI}
\end{equation}
where we have used the mean ($k=0$) relation $\langle \kappa I \rangle =   \langle j \rangle$ and defined a characteristic comoving distance photons travel, $R \equiv \langle I \rangle/(a \, b^{\kappa I}_{I,1} \langle \kappa I \rangle)$.   
We assume the perturbative limit $kR  \ll 1$, as our theory will break down on scales below the characteristic size of ionized structures. Thus, we can expand the denominator of eqn.~(\ref{eqn:deltaI}) to the desired order in $k R$.  Finally, we are interested in $\delta_\Gamma =  (4\pi)^{-1} \int d^2{\hat n} \delta_I$ -- the overdensity in total intensity -- and since $\delta_j$, $\delta_x$ and $\delta$ do not depend on ${\hat n}$, the angular average of eqn.~(\ref{eqn:deltaI}) to lowest order in $k R$ is
\begin{eqnarray}
b^{\kappa I}_{I,1} \widetilde{\delta_\Gamma} &=&  \left[1 - \frac{\left(k R\right)^2}{3} \right]  \left (\widetilde{\delta_j}(\bfk)  -b^{\kappa I}_{x, 1}\widetilde{\delta_{x}}(\bfk)  -b^{\kappa I}_{\delta, 1}\widetilde{\delta}(\bfk) \right) - \sum_{X,Y \in \delta, \delta_{x},\delta_\Gamma} b^{\kappa I}_{XY, 2}~ \widetilde{X} \star \widetilde{Y} . \nonumber
\end{eqnarray}
As justified in the previous section, we have dropped the ${\cal O}(k^4 \delta)$ terms as well as the ${\cal O} ( k^2 \delta^2)$ ones.  We will henceforth switch notation for the biases.  Using that $4\pi \kappa I =  \Gamma_p n_{\rm H} $, as both are just an expression for the number of absorptions, we will denote $b^{\kappa I}_X$ subsequently as $b^{ \Gamma n}_X$.  This notational change will make more apparent certain approximate cancelations below.  Finally, going to second order as desired for the 1-loop power spectrum, since terms that are locally cubic order can be absorbed in lower order terms at 1-loop, we can write 
\begin{eqnarray}
 \widetilde{\delta_\Gamma}^{(1+2)} =  \left[1 - \frac{\left(k R\right)^2}{3}\right]\,\widetilde{\delta_\Gamma^{*}}^{(1+2)}   - \sum_{X,Y \in  \delta^{(1)}, \delta_{x}^{(1)}, \delta_\Gamma^{*(1)}} \frac{b^{\Gamma n}_{XY, 2}}{ b^{\Gamma n}_{I,1}}~\widetilde{X}\star \widetilde{Y},
\label{eqn:deltaGamma2ndfinal}
\end{eqnarray}
where the $(1+2)$ superscript designates that terms only up to second order are included, and we have defined $ b^{\Gamma n}_{I,1} \widetilde{\delta_\Gamma^{*}} \equiv \widetilde{\delta_j} -b^{\Gamma n}_{x, 1}\widetilde{\delta_{x}} - b^{\Gamma n}_{\delta, 1}\widetilde{\delta} $. (Here the asterisk denotes lowest order in derivatives.)  

The other equation we consider aside from the radiative transfer equation is the equation for ionization balance:
 \begin{equation}
 \frac{dx_{\rm H}}{dt} = \overbrace{-\Gamma_{\rm p} x_{\rm H} + \Gamma_{\rm r} (1-x_{\rm H})}^{-[\Gamma x]},
 \label{eqn:xHdt}
 \end{equation}
 where $x_{\rm H}$ is the neutral hydrogen fraction, and $\Gamma_{\rm p}$  ($\Gamma_{\rm r}$) are the photoionization (recombination) rates.\footnote{The recombination rate is given by $\Gamma_{\rm r} \equiv \alpha  n_e$, where $\alpha$ is recombination coefficient and $n_e$ is the density of free electrons.}  We have defined the shorthand $-[\Gamma x]$ for the right hand side of the equation.  

To perturb around eqn.~(\ref{eqn:xHdt}) we write
\begin{equation}
[\Gamma x] = \langle [\Gamma x] \rangle \left(1 + b^{\Gamma x}_{I, 1,} \delta_\Gamma + b^{\Gamma x}_{x, 1} \delta_{x}  + b^{\Gamma x}_{\delta, 1} \delta +  \sum_{X,Y \in \delta, \delta_{x},\delta_\Gamma} b^{\Gamma x}_{XY, 2} [X Y - \langle XY \rangle] \right),
\end{equation}
using that $[\Gamma x] $ should be local in time and space in these fields (although see footnote \ref{footn}).
In the absence of recombinations ($ \Gamma_{\rm r} = 0$), we expect $b^{ \Gamma x}_X \approx b^{\Gamma n}_X$ (which remember is equal to $b^{\kappa I}_X$);  the bias of the regions that are being photoionized should be similar if volume-weighted or mass-weighted since reionization is the process in which the low density gas in the Universe is reionized as ionization fronts sweep over the Universe. Of course, equality cannot hold exactly.  

In Fourier space, equation~(\ref{eqn:xHdt}) becomes 
\begin{eqnarray}
\frac{d \widetilde{\delta_{x}}}{dt} &=& -\frac{d \log \bar x_{\rm H}}{dt} \widetilde{\delta_x} - \frac{\langle  [\Gamma x] \rangle}{\bar x_{\rm H}} \left( b^{\Gamma x}_{I, 1} \widetilde{\delta_\Gamma} + b^{\Gamma x}_{x, 1} \widetilde{\delta_{x}} + b^{\Gamma x}_{\delta, 1} \widetilde{\delta} +  \sum_{X,Y \in \delta^{(1)}, \delta_{x}^{(1)},\delta_\Gamma^{*(1)}} b^{\Gamma x}_{XY, 2} \widetilde{X} \star \widetilde{Y} \right), \nonumber
\label{eqn:dnHdteff}
\end{eqnarray}
where we denote the volume-averaged neutral fraction as $\bar x_{\rm H}$ for compactness of notation (and consistency with other sections).  Similarly, we define $\bar x_{\rm i} \equiv 1-\bar x_{\rm H}$ as the volume-averaged ionized fraction.

Plugging eqn.~(\ref{eqn:deltaGamma2ndfinal}) into this equation, 
 yields the master equation
\begin{eqnarray}
\frac{d \widetilde{\delta_{x}}}{dt} &=& - \left[ \frac{d \log \bar x_{\rm H}}{dt}+   \frac{\langle  [\Gamma x] \rangle}{\bar x_{\rm H}}  \left(b^{\Gamma x}_{x, 1}  - \frac{b^{\Gamma x}_{I, 1}}{b^{\Gamma n}_{I, 1}} b^{\Gamma n}_{x, 1}   \left[1 - \frac{\left(k R\right)^2}{3}\right] \right)  \right] \widetilde{\delta_{x}}  + {\cal I},
\label{eqn:dxHL}
\end{eqnarray}
where
\begin{eqnarray}
{\cal I} &= &  -\frac{\langle  [\Gamma x] \rangle}{\bar x_{\rm H}} \bigg[ \frac{b^{\Gamma x}_{I, 1} }{b^{\Gamma n}_{I, 1}} \left(\widetilde{\delta_j} - b^{\Gamma n}_{\delta, 1}\widetilde{\delta} \right)  \left[1 - \frac{\left(k R\right)^2}{3}\right] + b^{\Gamma x}_{\delta, 1} \widetilde{\delta} \\
 &&+  \sum_{X,Y \in \delta^{(1)}, \delta_x^{(1)},\delta_\Gamma^{*(1)}} \left( b^{\Gamma x}_{XY, 2} - b^{\Gamma n}_{XY, 2}  \frac{b^{\Gamma x}_{I, 1} }{b^{\Gamma n}_{I, 1}} \right) \widetilde{X} \star \widetilde{Y} \Bigg]. \nonumber
\end{eqnarray}

 Putting in explicitly the advection term with peculiar velocity field $\bfv$, the Green's function for the linear part of equation~(\ref{eqn:dxHL}) is
 \begin{eqnarray}
 G_k(t, t') &=&  \frac{\bar x_{\rm H}(t')}{\bar x_{\rm H}(t)} \exp\left[-\int_{t'}^t dt'' \left\{  \frac{\langle  [\Gamma x] \rangle}{\bar x_{\rm H}}  \left(b^{\Gamma x}_{x, 1}  - \frac{b^{\Gamma x}_{I, 1}}{b^{\Gamma n}_{I, 1}} b^{\Gamma n}_{x, 1}   \left[1 - \frac{\left(k R\right)^2}{3}\right] \right)  - i a^{-1} \bfv \cdot \bfk \right\} \right] \theta(t-t'),\nonumber
\end{eqnarray} 
  where all factors in the integrand depend on time.  We have assumed that $\delta_j$ does not depend on, e.g., $\delta_x$; such a dependence would arise from ionizing recombinations radiation and would add additional bias coefficients to the integrand.

The Universe starts out neutral before reionization without fluctuations in the neutral fraction $\widetilde{\delta_{x}}(t=0) = 0$.  Thus, the solution to eqn.~(\ref{eqn:dxHL}) for the overdensity in the neutral fraction to various orders in $\delta^{(1)}$ is
\begin{eqnarray}
\widetilde{\delta_{x}}^{(1)}(t) &=& \int^t_0 dt' G_k(t, t') {\cal I}^{(1)}(t'), \\
\widetilde{\delta_{x}}^{(2)}(t) &=&   \int^t_0 dt' G_k(t, t') {\cal I}^{(2)}(t'),\\
&...&
\end{eqnarray}
where $ {\cal I}^{(n)}(t')$ is evaluated with all terms that have $n$ powers of $\delta^{(1)}$.  For example, for $n=2$,  $[\delta^{(1)}]^2$ and $\delta^{(2)}$ can appear.

We do not need to evaluate explicitly the temporal integrals for $\delta_x$, since time dependences will be absorbed into effective coefficients, which we write as $b_{x,i(j)}$, and the perturbation theory will generate the expansion we found before (eqn.~\ref{eqn:general bias}).  Namely, for the terms needed at 1-loop in the power spectrum, if we drop the tidal terms and perform the time integrations
\begin{eqnarray}
 \tilde\delta_{x}^{(1)} &=&  b_{x,1(1)} \left(1- \frac{1}{3} R_{x, \rm eff}^2 k^2 \right) \tilde\delta^{(1)}, \label{eqn:d1}\\
  \tilde\delta_{x}^{(2)} &=&   b_{x,1(2)}\tilde \delta^{(2)} + b_{x,2(2)}\tilde \delta^{(1)} \star \tilde \delta^{(1)},\\
    \tilde\delta_{x}^{(3)} &=&  b_{x,1(3)}\tilde \delta^{(3)} + b_{x,2(3)} \tilde \delta^{(1)} \star\tilde  \delta^{(2)}, \label{eqn:d3}
\end{eqnarray}
where the configuration space fields are evaluated along Lagrangian trajectories (and, to be precise, displacements need to be expanded to yield terms with identical time dependences, resulting in the substitutions discussed after eqn.~\ref{eqn:general bias}).  We have dropped the terms that depend on the tidal field in equations~(\ref{eqn:d1})-(\ref{eqn:d3}), which are generated by the advection in the exponential.\footnote{The advection results in $ {\cal I}$ being evaluated at $\bfx -\Delta \bfpsi$, where we have defined the displacement vector $\Delta \bfpsi =  \int_{t'}^t dt'' i a(t'')^{-1} \bfv(t'')$.  Expanding, $ {\cal I} (\bfx - \Delta \bfpsi) = {\cal I}(\bfx) - \nabla_i {\cal I}(\bfx) \Delta \psi_i +  \frac{1}{2} \nabla_i \nabla_j {\cal I}(\bfx) \Delta \psi_i \Delta \psi_j +...$, creating tidal terms.} These terms are likely subdominant on the bubble scale because the distances ionization fronts travel are much larger than the matter displacements. Furthermore, on scales larger than the characteristic bubble size, these terms are still likely to be subdominant because the galaxy field sourcing reionization is highly biased (and their tidal bias is relatively small); the tidal terms generated by advection are suppressed relative to those tracing $\delta^2$ by factors of the source bias.

Let us work out a simple case to understand the predictions of our EPT of reionization. Many models find that the number of recombinations per ionization is just tens of percent, and so every ionizing photons is more or less balanced by an ionization.   In this limit, $\int_0^t dt' \langle  [\Gamma x] \rangle(t') = {\bar x}_{i}(t)$, not distinguishing here between mass-averaging and volume-averaging.  In addition, we have argued that 
roughly $b^{\Gamma n}_{X, 1} \approx b^{\Gamma x}_{X, 1}$ as the bias of ionizing regions should not depend strongly on whether we are weighting by density or by volume.  Let us assume equality, as this leads to large cancelations, simplifying our expressions. The Green's function simplifies to $G_k(t, t') = \bar x_{\rm H}(t')/\bar x_{\rm H}(t) \theta(t-t')$, ignoring the nonlocal and advection terms in the Green's function. However, we are more agnostic about the second order biases and the relation between $b^{\Gamma x}_{XY, 1}$ and $b^{\Gamma n}_{XY, 1}$, as these encode smaller effects.  Finally, let us take sources to be locally biased such that $\delta_j \approx b_{S, 1} \delta + b_{S, 2} \delta^2+...$, then the biases in equations~(\ref{eqn:d1})-(\ref{eqn:d3}) become
\begin{eqnarray}
 b_{x, 1(n)}(t)  &=& -\bar x_{\rm H} (t)^{-1} \int_0^t dt' \frac{d {\bar x}_{i}}{dt}(t') b_{S, 1}(t') \left(\frac{a(t')}{a(t)} \right)^n,  \label{eqn:b1simple} \\ 
b_{x,2(n)}(t) &= & -\bar x_{\rm H} (t)^{-1}\int_0^t dt' \frac{d {\bar x}_{i}}{dt}(t')  \left\{ \overbrace{b_{S, 2}(t')}^{\text{sources}}  + b^{\rm patchy}_{2}\right \}   \left(\frac{a(t'}{a(t)} \right)^n, \label{eqn:b2simple} \\
R_{x, \rm eff}^2 &=& \left[ {\bar x}_{\rm H}(t) b_{x,1(1)}(t) \right]^{-1} \int_0^t dt' \frac{d {\bar x}_{i}}{dt}(t') b_{S, 1}(t') R(t')^2 \left(\frac{a(t')}{a(t)} \right),\label{eqn:Reffsimple}
\end{eqnarray}
where $b^{\rm patchy}_{2}$ are the second order terms generated from the coupling of the radiation to opacity or between other overdensities (that derive from the $b^{\Gamma n}_{XY, 2}$ and $b^{\Gamma x}_{XY, 2}$).   Patchy reionization owes its patchiness to the large couplings between radiation and opacity and so we expect (and find) that these terms are large.  Finally, recombinations and the self shielding of dense regions should act to break the bias relations in this simple model. 

This section modeled the fluctuations in $x_{\rm H}$, a field we find in the next section is somewhat more perturbative than the very related 21cm signal.  We generalize our simple model to the 21cm signal in Section~\ref{ss:params}.



\section{testing the theory}
\label{sec:testing}
\label{sec:fit}

Our bias expansion and perturbation theory predicts shapes for the large-scale 21cm signal.  We could fit the power spectrum with these shapes and evaluate the goodness of fit, although the power spectrum is very broadband and so a good fit would never convincingly demonstrate that the theory is successful.  A more convincing validation fits our theory to every mode in a reionization simulation up to some maximum wavenumber.   Namely, to minimize ${\cal A}$ with respect to the parameters of the coefficients of our expansion, $\alpha_i$, where
\begin{eqnarray}
{\cal A} &\equiv& \sum_{k<k_{\rm max}} w_k P_{\rm err}(k, \alpha_i);\label{eqn:Perrmin}\\
&& P_{\rm err} = V^{-1} \left |\delta_{\rm err}(\bfk)\right |^2, ~~~\delta_{\rm err}(\bfk) \equiv  \delta_{X}(\bfk)  -  \sum_{\forall i} \alpha_i f_i(\bfk | \delta^{(1)}),
\label{eqn:Perr}
\end{eqnarray}
$V$ is the simulation volume, $w_k$ is a weighting function, $k_{\rm max}$ should be chosen such that the selected modes are perturbative, the $f_i(\bfk | \delta^{(1)})$ are the shape functions (with the summation in eqn.~\ref{eqn:Perr} running over all the desired shapes), and $\delta_X$ is the nonlinear ionization or 21cm field.  We use linear regression to minimize ${\cal A}$, 
 taking the weighting to be uniform such that $w_k=1$ and $k_{\rm max} = 0.2 \; h\,$Mpc$^{-1}$ or $0.4 \; h\,$Mpc$^{-1}$, which amounts to $176$ or $1248$ complex modes in our $130~ h^{-1}$Mpc simulations, many more than the $\leq7$ parameters that we aim to constrain.   Because the problem is so over-constrained, if a shape $f_i$ is not present, the fit is likely to prefer a negligible value for $\alpha_i$.  Missing shapes will manifest in our model underestimating the true signal power.
 
Our simulations' initial conditions provide $\delta^{(1)}$, and we use this to compute $\delta^{(2)}$ and $\delta^{(3)}$ following the method described in \citep{2016JCAP...03..007B}.  This method generates these Eulerian densities from the second and third order Lagrangian theory displacements, which are easier to compute. (We do not include the $k^2 \delta^{(1)}$ effective term that contributes to $\delta^{(3)}$ at 1-loop order and that is not in standard perturbation theory \cite[e.g.][]{carrasco12}.  This term is absorbed into our nonlocal biasing term.)\footnote{We have tested these displacements by calculating $P_{\rm err}$ for the nonlinear matter density field, finding orders of magnitude smaller values than for the 21cm field at relevant wavenumbers (with $P_{\rm err}/P_{\delta} <0.01$ at $k<0.8~h\,$Mpc$^{-1}$ and  $P_{\rm err}/P_{\delta} <10^{-4}$ at $k<0.3~h\,$Mpc$^{-1}$ at $z=8.3$), where $P_{\delta}$ is the power spectrum of the nonlinear matter overdensity.}  From the matter overdensity at various orders, the terms in the biasing expansion can be computed straightforwardly.

\subsection{fitting a minimal model to the 21cm signal}
Let us start off with a reduced model for the bias that does not separate terms that should be only unequal owing to their previous time dependence.  For example, we have set $b_{1(1)} = b_{1(2)}$ in eqn.~(\ref{eqn:general bias}).  This approximation is almost always made in bias expansions.  Section~\ref{ss:fullmodel} will show that the fits using the most general expansion are only marginally improved relative to this approximation.  

With this simplification, our bias expansion (or, equivalently, our perturbation theory) reduces to  
\begin{equation}
  \delta_{Y} = b_1 \left[1 - \frac{1}{3}R_{\rm eff}^2 k^2 \right] \widetilde{[\delta]} + b_2 \widetilde{[\delta^2]}, ~~~~~\text{\it (Minimal Model)}
\label{eqn:delta21}
\end{equation}
where either $Y=21$ (such that $\delta_{21} \equiv x_{\rm H} (1+\delta)$ is the 21cm `overdensity') or $Y=x$ (such that $\delta_{x}$ is the neutral fraction overdensity).  
  This {\it Minimal Model} has three bias coefficients ($b_1$, $R_{\rm eff}$, and $b_2$), and we have defined
\begin{eqnarray}
[\delta] &\equiv& \delta^{(1)} +\delta^{(2)} +\delta^{(3)},\\
\left[ \delta^2 \right] &\equiv & \left(\delta^{(1)} +\delta^{(2)}\right)^2 -\left(\delta^{(2)}\right)^2- \frac{68}{21} \sigma_L^2 \delta^{(1)},
\end{eqnarray} 
 which keeps only the terms that matter at one-loop order in the power spectrum (reminding you that $\delta^{(n)}$ is the $n^{\rm th}$-order matter overdensity in standard perturbation theory), and $\sigma_L$ is the variance of $\delta^{(1)}$.  
  The term with $\sigma_L$ renormalizes $[\delta^2]$ to be insensitive to ultraviolet modes to lowest order in derivatives \citep{2014JCAP...08..056A}, aside from shot noise.  This renormalization makes it so the fitted coefficients do not depend on the resolution of our calculations.\footnote{We find that using unnormalized quantities can result in much different bias coefficients.}  As a final simplification, we find that $\delta^2 = [\delta^{(1)}]^2$ yields slightly improved results with less sensitivity to the highest $k$ at which the field is smoothed.  We think this owes to the large shot noise in $\delta^2$. (Indeed, we find that even the full nonlinear density has less noise than $\delta^{(2)}$, which goes absolutely crazy on nonlinear scales; see e.g.~\cite{mcquinnwhite}.)  Finally, since we have not renormalized our expansion to terms that are higher order in derivatives, $R_{\rm eff}$ could have some sensitivity to the resolution of our calculations; we have checked that this sensitivity is weak. 
   
\begin{figure}
\begin{center}
\epsfig{file=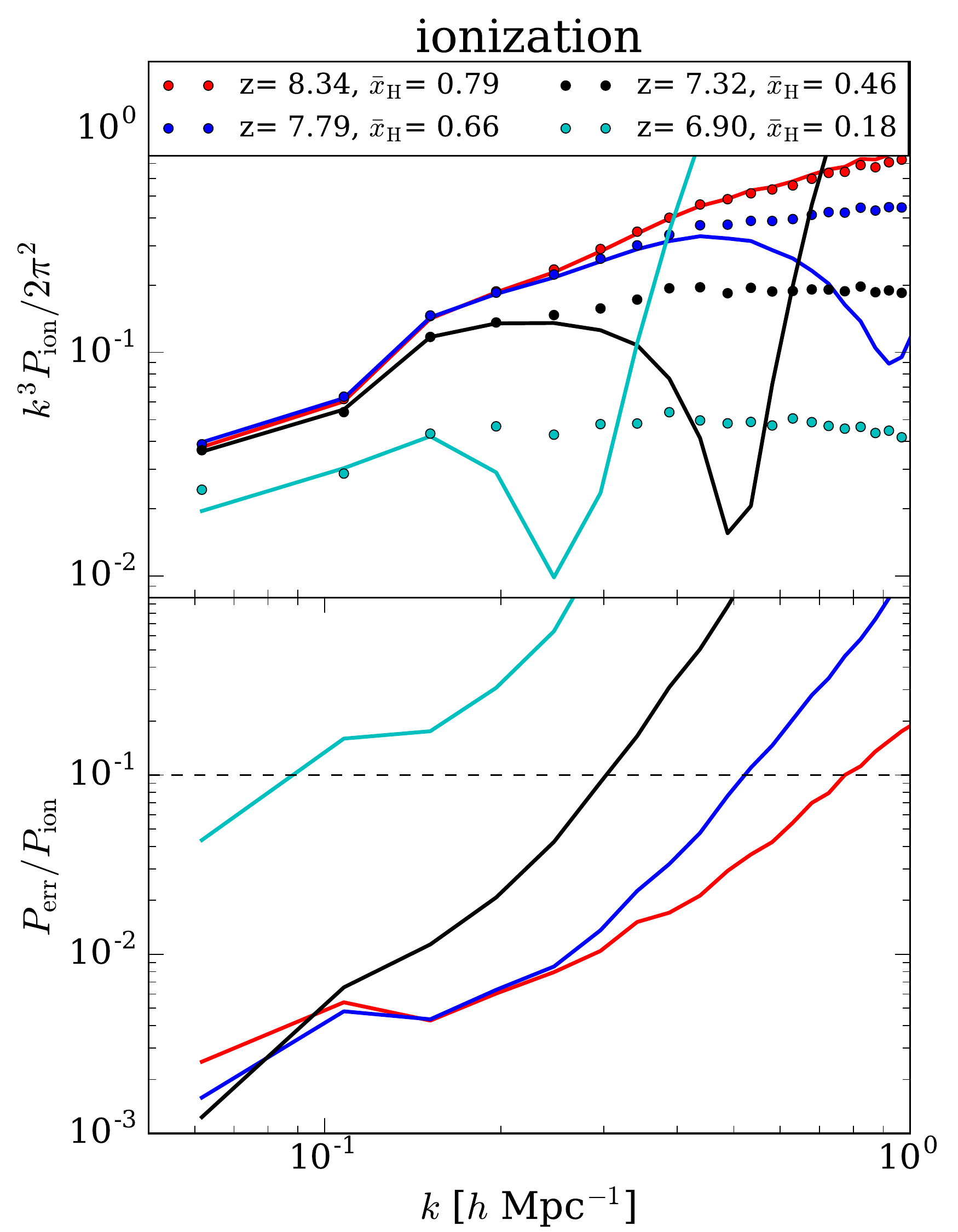, width=7.6cm}
\epsfig{file=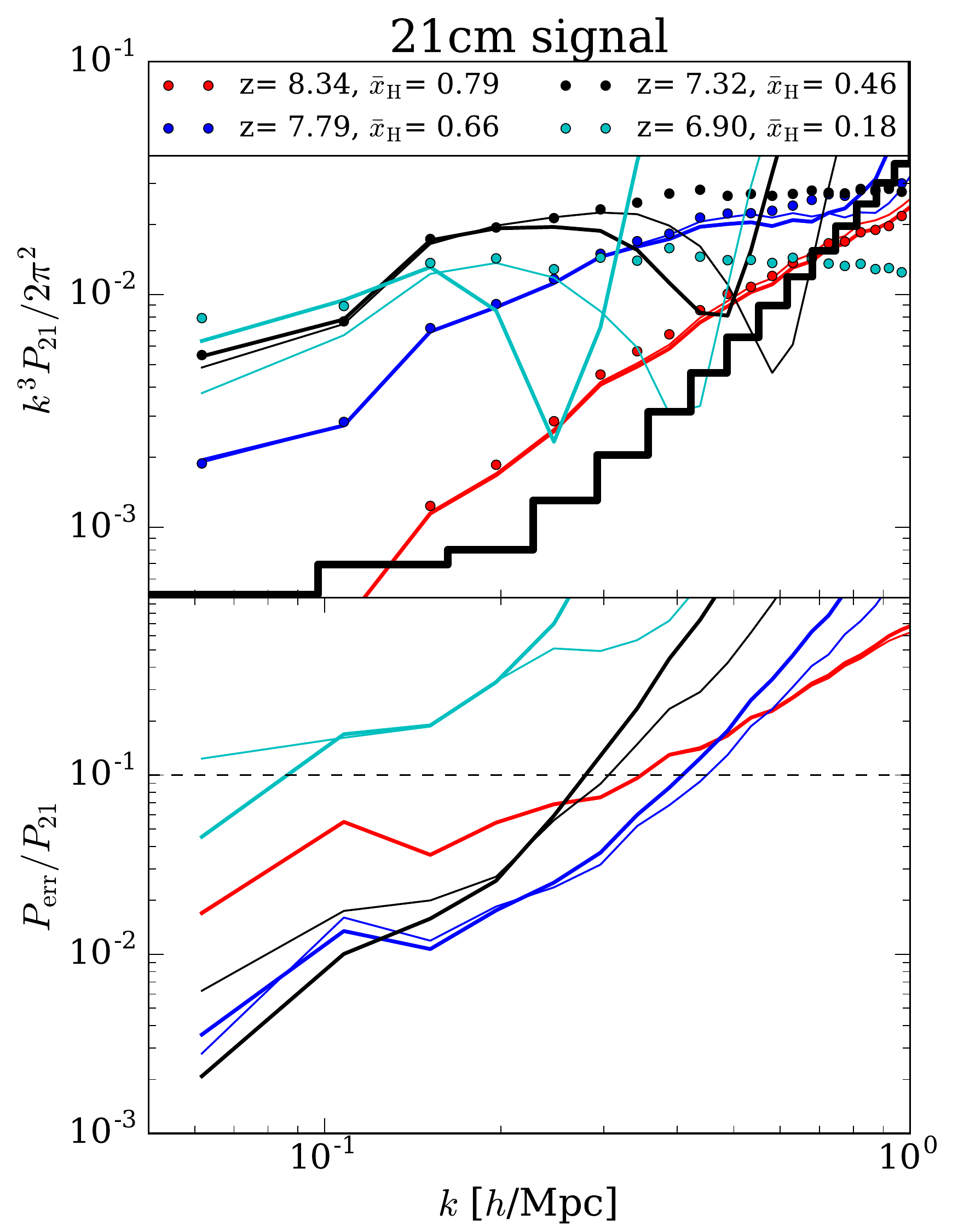, width=7.6cm}
\end{center}
\caption{Fit of Fiducial reionization simulation to the $3$-parameter {\it Minimal Model} given by eqn.~(\ref{eqn:delta21}).  The lefthand panels consider the ionization fraction overdensity, and the righthand ones consider the 21cm `overdensity' $\delta_{21} \equiv x_{\rm H} (1+\delta)$. The upper panels feature the power spectrum of these quantities, with the filled dots showing this signal in our Fiducial simulation and with the solid curves showing the best-fit {\it Minimal Model}.  Remember that this model fits for the bias coefficients $b_1, b_2$ and $R_{\rm eff}$.  The lower panels feature the power spectrum of the model error (eqn.~\ref{eqn:Perr}), divided by the simulation power spectrum.  The thick curves fit the model to modes as large as $k_{\rm max} = 0.2~h\,$Mpc$^{-1}$, and the thin fit to modes as large as $k_{\rm max} = 0.4~h\,$Mpc$^{-1}$.  The black solid histogram in the upper-right panel is the forecasted $z=8$ sensitivity of HERA.  The horizontal dashed line indicates where the error power spectrum is $10\%$ of the simulation power spectrum, a benchmark referred to in the text.
 \label{fig:Pkanderrors}}
\end{figure}

The upper panels in Figure \ref{fig:Pkanderrors} show the power spectrum of the Fiducial simulation (filled dots) as well as the best-fit {\it Minimal Model} (solid curves) for four snapshots.  The lower panels show for the same four snapshots the power spectrum of the model error (eqn.~\ref{eqn:Perr}) divided by power spectra of the Fiducial simulation.  The thick curves fit the {\it Minimal Model} to wavenumbers of $k_{\rm max} = 0.2~h\,$Mpc$^{-1}$, and the thin to $k_{\rm max} = 0.4~h\,$Mpc$^{-1}$.  The thick and thin curves should agree if both are fit to scales that are well described by the model.

The lefthand panels in Figure \ref{fig:Pkanderrors} show the ionization fraction power spectrum, $P_{\rm ion}$.  The model fits to the snapshots with $\bar x_{\rm H} = 0.79$, $0.66$, and $0.46$ satisfy the goodness condition $P_{\rm err}/P_{\rm ion} < 0.1$ at wavenumbers of $k < 0.8$, $0.5$, and $0.3 ~h\,$Mpc$^{-1}$, respectively.  The errors are larger for the $x_{\rm H} = 0.18$ snapshot -- the plotted snapshot for which the ionized bubbles are largest and the ionization field the least perturbative --, with $P_{\rm ion}/P_{\rm err} < 0.1$ only at $k \lesssim 0.1 ~h\,$Mpc$^{-1}$.

The righthand panels in Figure \ref{fig:Pkanderrors} show the power spectrum of the 21cm signal, $\delta_{21} \equiv x_{\rm H} (1+\delta)$. (More precisely, $x_{\rm H} (1+\delta)$ is the 21cm brightness temperature divided by the mean 21cm brightness temperature for a fully neutral universe.)   We find the 21cm field to be slightly less perturbative than the ionization field, with $P_{\rm err}/P_{21} < 0.1$ only at $k < 0.25-0.4~h\,$Mpc$^{-1}$ for the $\bar x_{\rm H} = 0.79$, $0.66$, and $0.46$ snapshots.  
  The solid black histogram in the upper-right panel shows HERA forecasts for the error power spectrum at $z= 8$ for band-powers of $\Delta k =0.06\;h\,$Mpc$^{-1}$ using the `moderate' assumptions for sensitivity loss owing to foreground removal of \citep{2014ApJ...782...66P}, updated to recent HERA specifications (provided by J.~Pober).  This histogram suggests that, with the exception of the $x_{\rm H} = 0.18$ snapshot, the {\it Minimal Model} is successful for a significant fraction of the observable wavenumber range.  

\begin{figure}
\begin{center}
\epsfig{file=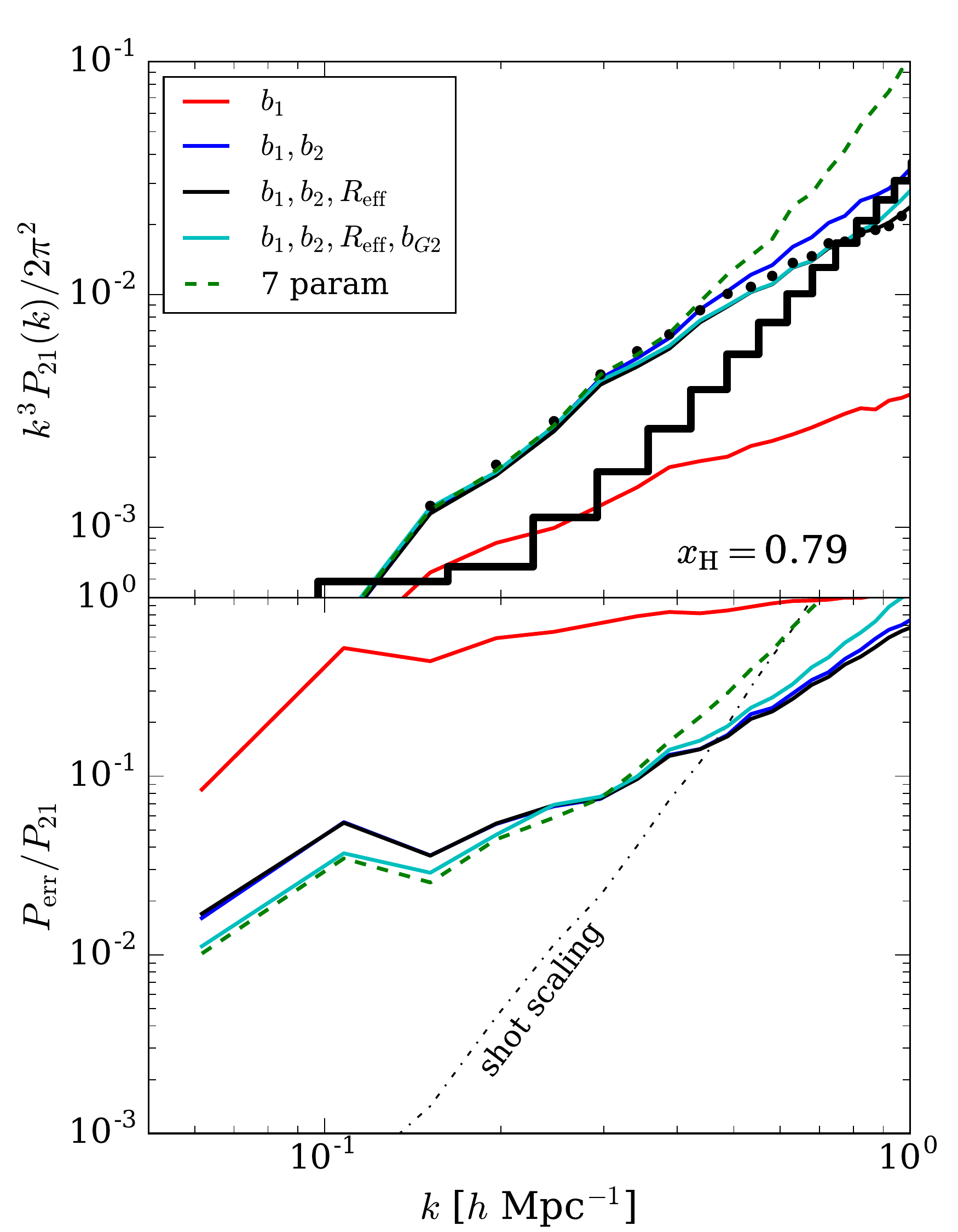, width=7.6cm}
\epsfig{file=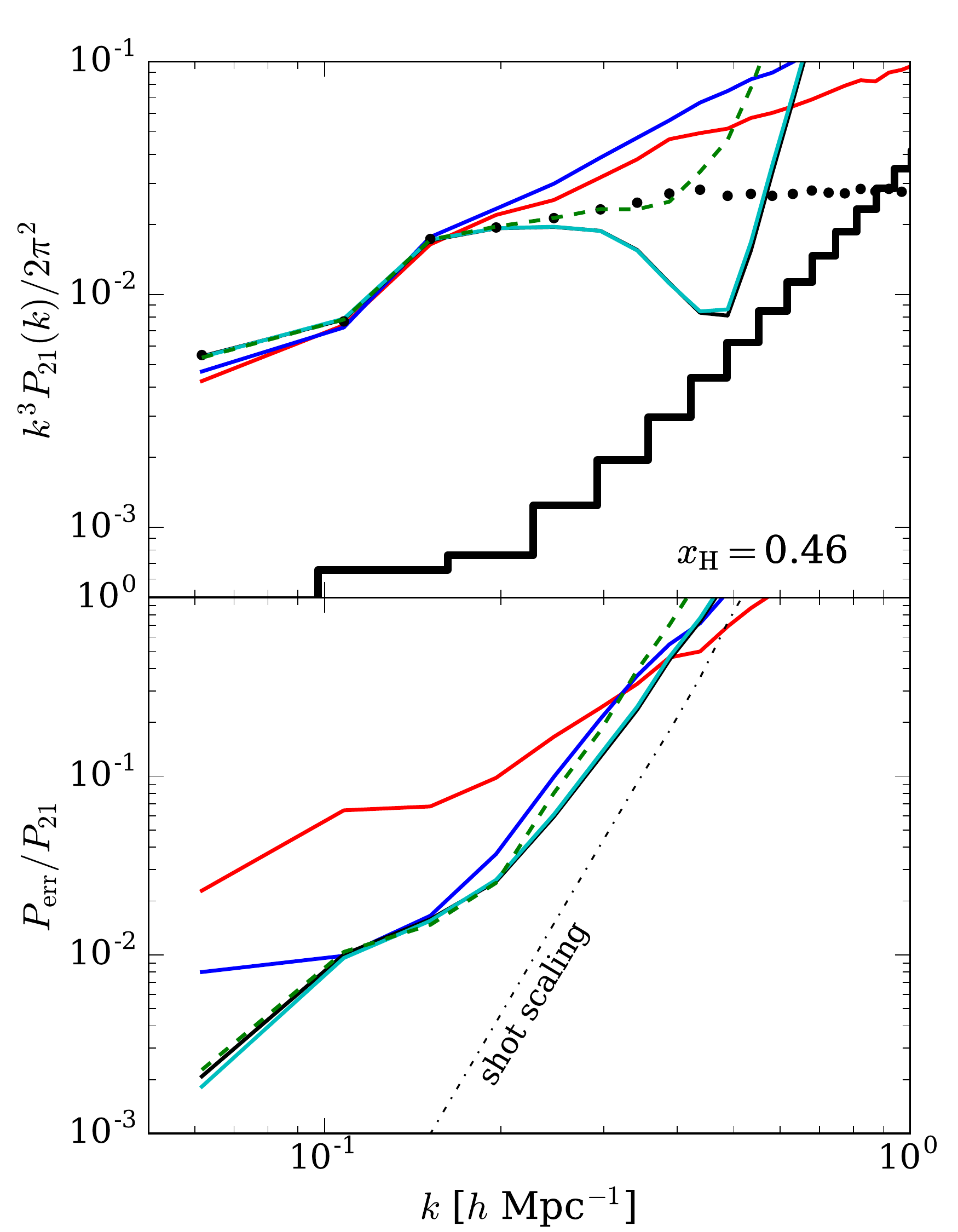, width=7.6cm}
\end{center}
\caption{Fits using different combinations of bias parameters, with the parameters specified in the legend. The parameters $b_1$, $b_2$ and $R_{\rm eff}$ comprise the {\it Minimal Model} (eqn.~\ref{eqn:delta21}), and $b_{G2}$ is the quadratic tidal bias (see \S~\ref{sec:bias}). The green dashed curves show the results of our more general seven parameter fit (Section~\ref{ss:fullmodel}).   The black dot-dashed curves in the lower panels show the expected scaling if $P_{\rm err}$ were white; flatter $P_{\rm err}/P_{21}$ than the dot-dashed scaling suggest non-perturbative effects.
 \label{fig:withnumparameters}}
\end{figure}   

Figure~\ref{fig:withnumparameters} shows how the fit to $\delta_{21}$ improves as we increase the number of parameters.  The fits are improved dramatically by including $b_2$ in addition to the linear bias coefficient $b_1$, especially for the $\bar x_{\rm H} = 0.79$ case shown in the lefthand panels (because $b_1$ is near its zero crossing, as discussed in \S~
\ref{ss:params}).   Fitting for the effective bubble size, $R_{\rm eff}^2$, only impacts the fit somewhat and, then, generally only at the largest wavenumbers to which this expansion applies.  The bubble size parameter becomes more important at smaller $\bar x_{\rm H}$ than considered in this figure, with this parameter being the only one able to impart the flattening in the power spectrum that has been thought to indicate a broad spectrum of bubbles.  Fitting for the $2^{\rm nd}$--order tidal term (denoted by $b_{G2}$ in the figure) does not improve the fits appreciably, as surmised in Section~\ref{sec:PT}.  The dot-dashed curves in the lower panels of Figure~\ref{fig:withnumparameters} are the expected scaling for shot noise ($P_{\rm err} =$constant).  Shot noise may explain the high wavenumber behavior of the residuals for $\bar x_{\rm H}=0.46$, but there is a significant non-shot structure in $P_{\rm err}/P_{21}$ at lower wavenumbers.  Most of this residual is not removed by our most general expansion that allows for different time dependences (which we will discuss in Section~\ref{ss:fullmodel}).  Since higher-order terms than those accounted for in our expansion should scale more strongly with $k$ than shot, the residuals must owe to non-perturbative effects.  The model error from non-perturbative effects almost certainly arises from the largest (and rarest) ionized regions, as our expansion in $kR$ breaks down for $k \gtrsim R^{-1}$.\footnote{{\refereeedit Perhaps surprisingly, the fractional contribution of such non-perturbative effects is larger for the $\bar x_{\rm H} = 0.79$ snapshot compared to $\bar x_{\rm H}=0.46$ one, even though the bubbles are smaller in the former case.  However, at $k\approx 0.2~h$Mpc$^{-1}$ the $\bar x_{\rm H} = 0.79$ case has a factor of ten less power in $P_{21}$ than $\bar x_{\rm H} = 0.46$ case, and yet $P_{\rm err}/P_{21}$ is only a factor of $\sim2$ smaller (see Fig.~\ref{fig:withnumparameters}).  Therefore, the absolute size of the non-perturbative residuals ($P_{\rm err}$) are still much smaller in the $\bar x_{\rm H} = 0.79$ case and appear to grow monotonically with time as expected.}}

\begin{figure}
\begin{center}
\epsfig{file=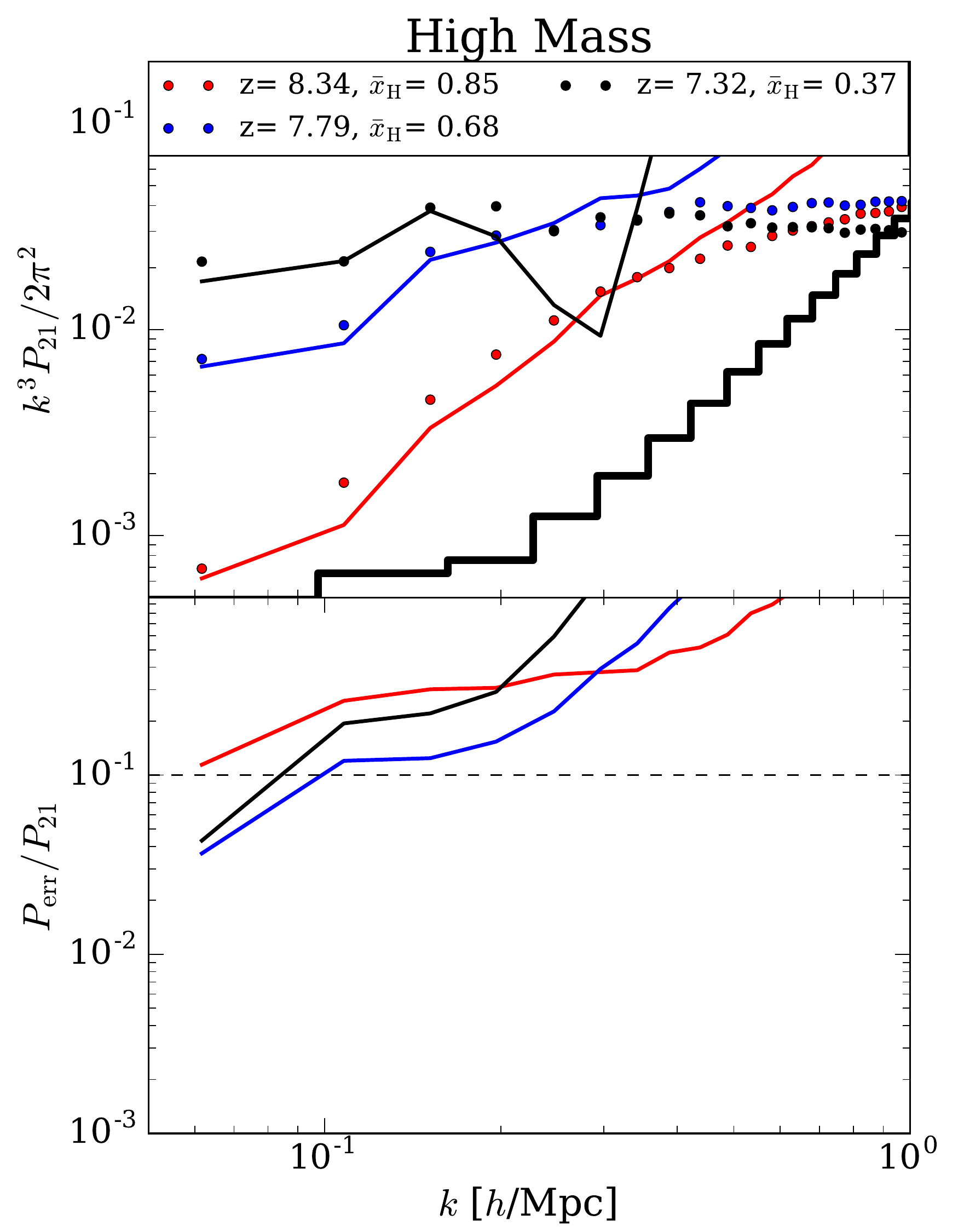, width=7.6cm}
\epsfig{file=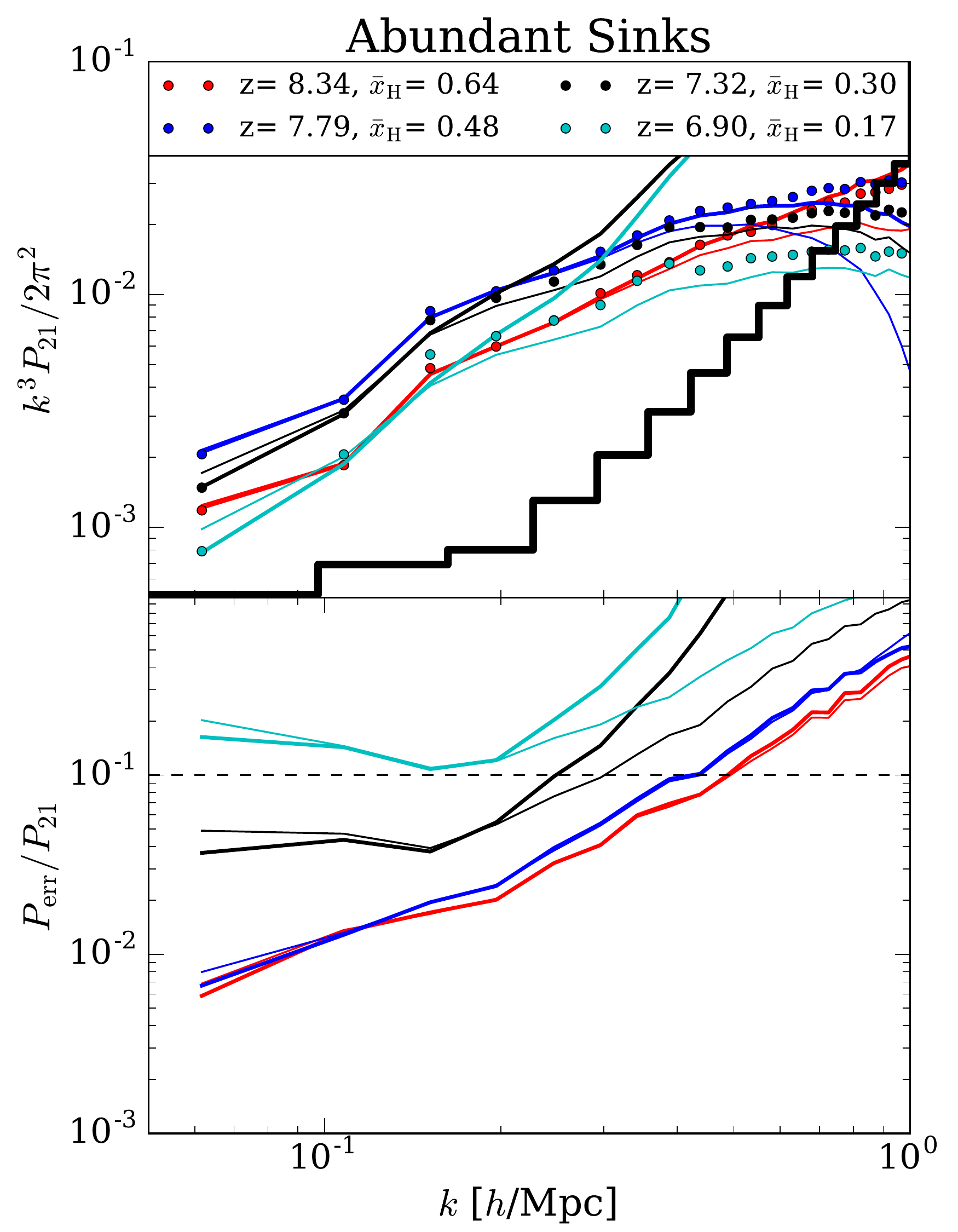, width=7.6cm}
\end{center}
\caption{The same as Fig.~\ref{fig:Pkanderrors} except showing the 21cm signal for two other reionization simulations plus the best-fit {\it Minimal Model} (which fits for $b_1$, $b_2$, and $R_{\rm eff}$).  The lefthand panels shows our High Mass simulation, which results in the largest bubbles of the three simulations and, hence, is the least perturbative.  The righthand panels show our Abundant Sinks simulation, which results in the smallest bubbles.  We only include the $k_{\rm max} = 0.2 h~$Mpc$^{-1}$ (thick) curves, excluding the $k_{\rm max} = 0.4 h~$Mpc$^{-1}$ (thin) curves, for the High Mass case as the perturbative solution is breaking down at smaller wavenumbers.  
 \label{fig:othermodels}}
\end{figure}    

The lefthand panels in Figure~\ref{fig:othermodels} show how well the {\it Minimal Model} describes the High Mass simulation. We note that the {\it Minimal Model} fit generally underestimates the power, more than for the Fiducial model, which is expected because the high-mass simulation results in the largest bubbles of the three simulations and, hence, should have non-peturbative shapes not described by our theory.   Still, the signal trends are largely described by the {\it Minimal Model}:  The non-perturbative component results in $\sim 10\%$ errors in $P_{21}$ at $k\lesssim 0.2-0.3~h\;$Mpc$^{-1}$.  The High Mass simulation should have a larger shot noise term than the other simulations, which would contribute to $P_{\rm err}$; however, the spectrum of the residuals do not look consistent with shot noise.  Despite the larger error, the level of accuracy is likely comparable to that of semi-analytic reionization models at these wavenumbers.  (If the {\it Minimal Model} were fit to $P_{21}$ rather than $\tilde \delta_{21}$, the fit would be more accurate because then the fitted terms can compensate for missing ones.)  

The righthand panels Figure~\ref{fig:othermodels} show how well the {\it Minimal Model} describes our Abundant Sinks simulation.  In this simulation, the photon mean free path is limited by a subgrid recipe for absorptions, resulting in smaller bubbles than in the Fiducial simulation, especially during the latter half of reionization.  Relative to the other simulations, $P_{21}$ at $k\sim 0.1\;h\,$Mpc$^{-1}$ can be lower by an order of magnitude at fixed $\bar x_{\rm H}$, suggesting that the 21cm signal from this simulation is more perturbative.  Indeed, the {\it Minimal Model} is somewhat more successful at describing the Abundant Sinks simulation compared to the Fiducial simulation.

\subsection{interpreting the bias parameters in the {\it Minimal Model} fits}
\label{ss:params}

\begin{figure}
\begin{center}
\epsfig{file=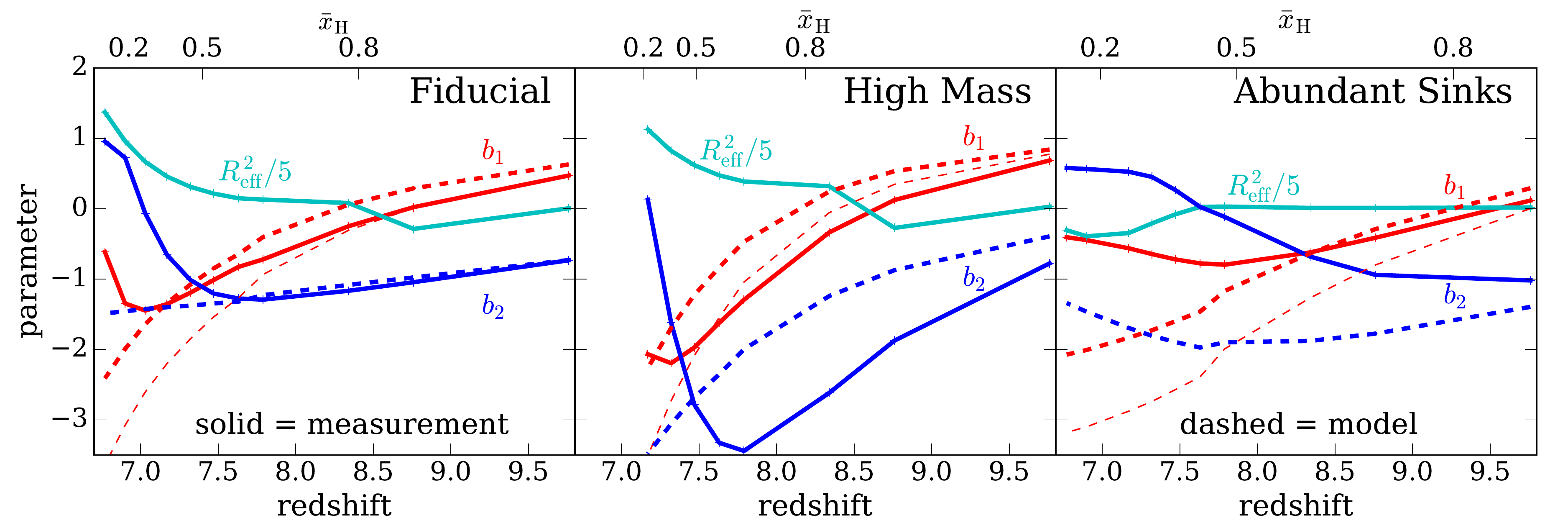, width=15cm}
\end{center}
\caption{Best-fit bias parameters in the {\it Minimal Model} to $\delta_{21} \equiv x_{\rm H} (1+\delta)$ as a function of the redshift (lower $x$-axis) or $\bar x_{\rm H}$ (upper $x$-axis).   Each panel features one of our three reionization simulations, and the fits are done using modes with $k<0.2~h\,$Mpc$^{-1}$.  The thick dashed curves show the predictions of a simple theory for $b_1$ and $b_2$ in which the ionization traces the sources.  The thin red dashed curves are the same models for $b_1$ as the thick red dashed curves except that the linear source bias $b_{S, 1}$ is inflated by $50\%$.
 \label{fig:fittingparameters}}
\end{figure}    

Figure~\ref{fig:fittingparameters} shows the resulting parameters from the {\it Minimal Model} fits to the three reionization simulations.  In all the simulations, the linear bias, $b_1$, starts off positive and eventually becomes negative, whereas $b_2$ starts off negative, becomes more negative, and then increases quickly towards the end of reionization.  The zero crossing of $b_1$ occurs where the slope of $P_{21}$ is the steepest, as $P_{21}$ becomes dominated by the power spectrum of $[\delta^2]$.  

A simple theory in which the ionized bubbles trace the locations of the sources in proportion to their ionizing luminosity predicts $b_1 = \bar x_{\rm H} - \bar x_i b_{S, 1}$ and $b_2 =  - \bar x_i b_{S, 2}$, where the source overdensity is assumed to follow $\delta_S = b_{S, 1} \delta + b_{S, 2} \delta^2$.  This model is derived by assuming only that the ionization overdensity traces that of the sources such that $\delta_{n_i} = \delta_S +\delta$ (roughly the ionized fraction overdensity equal to $\delta_S$), where $n_i$ is the density of ionized gas and the $+\delta$ term is needed to reproduce the zero bias limit.  Using that $\bar n \delta = \bar n_{\rm H} \delta_{n_{\rm H}} +  \bar n_i \delta_{n_i}$, $n= n_i + n_{\rm H}$, it follows that $\delta_{21} = \bar{x}_{\rm H} \delta_{n_{\rm H}} =  \bar{x}_{\rm H}\delta  - \bar{x}_i \delta_S$.  This simple linear bias theory ignores recombinations, which, if substantial, act to reduce $b_1$; recombinations slow growth in the largest and, hence, most biased regions \citep[see framed inset for more discussion]{furlanetto05,mcquinn07}.
 \begin{framed} 
\paragraph{On the validity of tying $b_1$ to source clustering:} Our simple model that derives $b_1$ assuming the ionization traces the sources should hold in the absence of recombinations (and if we can ignore light travel delays, which is likely a good approximation).  As $b_1$ applies on any perturbative scale, we can choose a scale much larger than that of the bubbles.  Then, without recombinations every ionizing photon results in an ionization.  A region with $X$ times more sources will have $X$ times more ionization.  So, in this limit, the linear ionization bias must equal the source bias. $\square$
\end{framed}
  The simple theory's predictions for $b_1$ and $b_2$ are shown by the thick dashed curves in Figure~\ref{fig:fittingparameters}, using Extended Press-Schechter theory to compute the sources' Eulerian biases \citep{2001MNRAS.323....1S, cooray02}.  The thin red dashed curve is the same model for $b_1$ except that $b_{S,1}$ has been increased by $50\%$, which provides a better match.  We note that sources in $<8\times10^9\Msun$ halos that are unresolved in these simulations were put down with an algorithm that is based on Extended Press-Schechter theory \citep{mcquinn11}, following \citep{sheth99}.  Because the source fields used for the simulations no longer exist, we do not exactly know the source bias in the simulations and, hence, limit our discussion to the qualitative trends.  We also note that the somewhat more complex model given by equation~(\ref{eqn:b1simple}) in which $b_{1,x}$ equals a time integral over $b_{1,S}(z)$ yields essentially the same prediction as the simple model above (using that the ionization overdensity is $ -\frac{\bar x_{i}}{\bar x_{\rm H}} \delta_x$; this model reduces to our simple expression $\delta_{21} =  \bar{x}_{\rm H}\delta  - \bar{x}_i \delta_S$ if all ionization occurs at the redshift of interest).


This simple theory's $b_1$ matches the trends in the simulations' $b_1$, even towards the very end of reionization in the Fiducial and High Mass simulations -- a surprise since the 21cm signal goes to zero (whereas this theory predicts $b_1 \rightarrow -b_{S, 1} \neq 0$).  This occurs largely because the 21cm signal has yet to decline at low wavenumbers in most of our simulations by the last snapshot that was stored (as most of our simulations only reach $\bar x_H \approx 0.15$).  Recombinations should eventually cause $b_1$ to deviate from our simple expression (as the bubbles overlap and the most biased sources' photons become preferentially absorbed within dense regions in the bubbles, decreasing $b_1$).  Indeed, $b_1$ in the Abundant Sinks simulation differs from the simple theory at much earlier times than the other simulations, once $\bar x_{\rm H}<0.5$.  

Let us now consider the $[\delta^2]$ bias coefficient, $b_2$. The comparison between the fitted $b_2$ and the simple biasing theory (see Fig.~\ref{fig:fittingparameters}) suggests that $b_2$ is set by the source clustering for the first half of reionization such that $b_2 \approx - \bar x_i b_{S, 2}$, again noting the roughness of our extended Press Schechter estimate for $b_2$. Once $\bar x_{\rm H} \lesssim 0.6$, the patchy contribution to this coefficient -- from the coupling of fluctuations in the neutral gas to the radiative transfer -- begin to kick in and eventually dominate. The result of these terms is to cause $b_2$ to increase from negative values and eventually become positive.  The values of $|b_2|$ during most of reionization are larger than the values of $|b_1|$ by at least a factor of few (and their contribution to $P_{21}$ scales quadratically in each bias as the $b_1$ and $b_2$ tracing terms are orthogonal).

Our final parameter in the {\it Minimal Model} is the square of the effective bubble size, $R_{\rm eff}^2$.  This parameter grows from a small value to greater than $10~$Mpc$^2$ in the Fiducial simulations, and evolves similarly in the High Mass run.  This parameter is most responsible for the characteristic flattening of the signal that is observed.  We note that the resulting $R_{\rm eff} \sim 3~$Mpc is a smaller characteristic scale than picked out by eye or in excursion set models \citep{furlanetto04a, zahn06};  its smallness is also a reason why perturbation theory is successful at the wavenumbers of interest, as our perturbative expansion should break down once $k R_{\rm eff} /\sqrt{3} \gtrsim 1$.  Some intuition for the smallness of $R_{\rm eff}^2$ may be gleaned from our expression for this parameter given by equation~(\ref{eqn:Reffsimple}):  The effective bubble size is suppressed by factors of (the likely large) source bias (since $R \equiv \langle I \rangle/[a  b^{\kappa I}_{I,1} \langle \kappa I \rangle]$) and weighted over previous times, when the mean free path, $\kappa^{-1}$, was smaller.  Physically, while slices through the ionization field show large bubbles (e.g.~Fig.~\ref{fig:xill}), there is also structure on smaller scales, especially in the neutral regions, and $R_{\rm eff}$ must account for these as well.\footnote{In the fits to the Abundant Sinks simulation, $R_{\rm eff}^2$ becomes slightly negative at late times because this parameter is not well constrained: In the fit for $k_{\rm max} = 0.4~h\,$Mpc$^{-1}$ rather than $k_{\rm max} = 0.2~h\,$Mpc$^{-1}$, $R_{\rm eff}^2$ is essentially zero at all times in this model [with other parameters unaffected].  In addition, in all three simulations, $R_{\rm eff}^2$ is not well constrained at early times when the bubbles are smallest.}

\subsubsection{fitting the minimal model to $P_{21}$}
In addition to fitting individual modes directly, we have also fit the shapes in the {\it Minimal Model} to the simulated 21cm power spectrum.  Fortunately, $[\delta^{(1)}]^2$ is orthogonal to the one loop $\delta$, and so the power spectrum shapes we must fit are the linear power spectrum $P_L$,  its convolution $P_L\star P_L$ (which is the power spectrum of $\delta^2$), and $k^2 P_L$.  It is only the last term that can take negative values since the $b_1$ and $b_2$ bias coefficients appear in quadrature in the first terms. 
 (We also have investigated including a shot noise term in the fit, which for the High Mass simulation modestly improves the fit.)  In contrast to fitting individual modes, where the parameters are over-constrained, for fitting to the power spectrum we only use a few broad band-powers to mimic what would be done on the actual observations.  In particular, we fit the $n$ lowest wavenumber band-power bins shown in Figures~\ref{fig:Pkanderrors} and \ref{fig:othermodels}, assuming an error bar in each bin that is proportional to the signal.  These fits often look visually better than the fits to the modes, since the latter are less willing to use an incorrect shape to describe the signal.

  We find that the linear bias is generally well constrained even with $S/N$ of just a couple in each band-power bin and $n$ greater than or equal to the number of fitted parameters.  However, we find that $b_2$ is hard to constrain even with more band power bins, with the exception being when $b_1$ is small as occurs at $\bar x_{\rm H} \approx 0.1$.  (A bispectrum measurement may better isolate this term.)  We find that the requirements for constraining $R_{\rm eff}^2$ are even more stringent.   A full analysis is rather involved, as the number of free parameters and the maximum fitted wavenumber should be motivated by the amplitude and shape of the measured $P_{21}$.  We leave such an analysis to future work.

\subsection{fitting all 1-loop terms to $\delta_{21}$}
\label{ss:fullmodel}

So far we have considered a simpler parametrization for the 21cm signal.  The most general biasing expansion, outlined in Sections~\ref{sec:bias} and \ref{sec:PT}, should improve the fits.  To understand the magnitude of the improvement, we have fit the seven coefficients for the most general 1-loop expansion, given by equation~\ref{eqn:general bias}.  (There are eight coefficients in this expression;  to reduce this to seven, we do not separate the ${\cal G}_2$ tidal term by components that arise from time dependences, assuming $b_{G2(2)} =b_{G2(3)}$.)  The green dashed curve in each panel of Figure~\ref{fig:withnumparameters} are this best-fit seven parameter model, fitting to $k_{\rm max} = 0.2\;h\,$Mpc$^{-1}$ like the other model curves in this figure.  Comparing to other model curves, which fit fewer parameters, we find no significant improvement in the seven parameter model's goodness of fit.  
 
 We think that one reason fitting these extra terms does not improve the goodness of fit is not that, for example, $b_{1(1)} =b_{1(2)} =b_{1(3)}$, but instead because linear theory provides a good approximation for the matter density field on the wavenumbers where our expansion is successful.  Stated in another way, the nonlinear terms from patchy reionization and not from the density are most important for the shaping mildly nonlinear scales in the 21cm signal.   Figure~\ref{fig:CCC} supports this assertion, showing that the absolute value of the cross-correlation coefficient between the linear and nonlinear matter overdensity is closer to unity than the cross-correlation coefficient between the linear matter overdensity and 21cm signal.  This result may elucidate why semi-analytic algorithms, where the ionization only depends on the coeval density, successfully describe the signal.  
 
 However, another reason why the more complex expansion is not more successful is that each term in this expansion has a shot component (as, e.g., squaring $\delta$ will result in a white term), and the wavenumber dependences  of the higher order terms included in the expansion are less distinct from these terms shot noise.  This shot noise makes these higher-order terms more noisy templates when minimizing $\delta_{\rm err}$.  
   


\subsection{Could some future perturbative model do better?}

\begin{figure}
\begin{center}
\epsfig{file=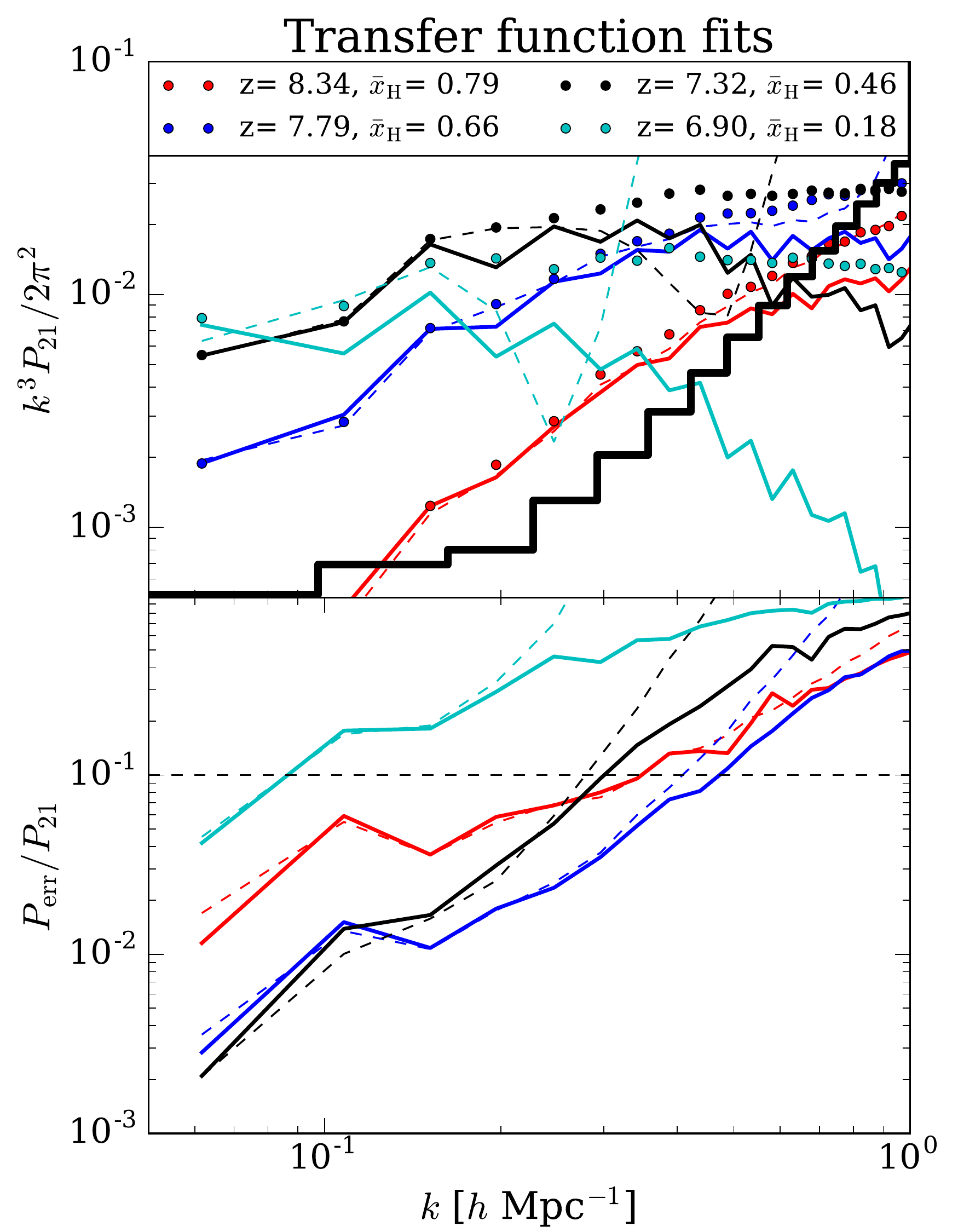, width=7.6cm}
\end{center}
\caption{Similar to Fig.~\ref{fig:Pkanderrors} but showing the improvement of a transfer function model in which $\delta_{21} = b_1(k) \widetilde{[\delta]} + b_2(k) \widetilde{[\delta^2]}$ over the {\it Minimal Model} (in which $b_1$ and $b_2$ are constants and there is a term that scales as $k^2 \delta^{(1)}$).  The solid curves are the best-fit {\it Transfer Function Model}, and the dashed curves are the best-fit {\it Minimal Model}.  The reduction in $P_{\rm err}$ for the {\it Transfer Function Model} is significant, but only once $P_{\rm err} > 0.1$ such that perturbation theory is already failing.
 \label{fig:tfuncmodel}}
\end{figure} 

The cross correlation coefficients between the 21cm signal and linear density field generally have $|r| >0.8$ to a factor of two or so higher wavenumbers than to where our perturbative model is able to reproduce the signal with $P_{\rm err} \lesssim 0.1$ (see Fig.~\ref{fig:CCC}).  This suggests to us that there might be a better ``perturbation theory'' in which nonperturbative physics (or possibly resummations of higher order perturbative terms) can be encapsulated in a theory that traces $\delta^{(1)}$ with a wavenumber-dependent transfer function. We note that in the perturbation theory of the cosmological matter field, such an approach has been found to improve the success of the 1-loop theory \citep{tassev13, 2016JCAP...03..007B}, which occurs because higher-order terms often largely trace lower order terms.  

To test this possibility, we investigate a {\it Transfer Function Model} that allows the $b_1$ and $b_2$ coefficients of the {\it Minimal Model} to vary with wavenumber, fitting them in wavenumber bins of $\Delta k = 0.1~h\,$Mpc$^{-1}$ to the Fiducial simulation.  Figure~\ref{fig:tfuncmodel} shows the results of this exercise.  The dashed curves are the best-fit {\it Minimal Model} for $k_{\rm max} = 0.2 ~h\;$Mpc$^{-1}$, and the solid curves are the best-fit {\it Transfer Function Model}.   In these transfer function models, the value of $b_1(k)$ changes by a factor of $\{0.72, 0.66, 0.42, 0.14\}$ between $k=0.05~h\,$Mpc$^{-1}$ and $k=0.55~h\,$Mpc$^{-1}$ for the $\bar x_{\rm H} = \{0.79, 0.66, 0.46, 0.18\}$ shown in this figure.  While both models produce $P_{\rm err}/P_{21}=0.1$ at similar wavenumbers, the errors grows more slowly in the {\it Transfer Function Model} fits at higher wavenumbers: In all four snapshots shown, the transfer function fits have $P_{\rm err}/P_{21} < 1$ at $k=1~h\,$Mpc$^{-1}$, significantly smaller $P_{\rm err}/P_{21} $ than the {\it Minimal Model} fits.  
 These differences explain why the cross correlation coefficients are still significant at $k\sim 1~h\;$Mpc$^{-1}$.  Yet, on scales where these models are relatively accurate ($P_{\rm err}/P_{21}<0.1$), the transfer function model provides minimal improvement.

\section{conclusions}

This paper developed a bias expansion and, equivalently, an effective perturbation theory to describe spatial fluctuations both in the ionization field and the 21cm radiation field during cosmological reionization.  We compared these models to radiative transfer simulations of reionization, showing that the simulated signal is perturbative  over much of the observationally-relevant wavenumber range.  These models predict that the large-scale 21cm signal must have certain shapes, which we argued can be simplified to 
\begin{equation}
 \delta_{21} = b_1 \left(1 - \frac{1}{3}R_{\rm eff}^2 k^2 \right) \widetilde{\delta} + b_2 \; \widetilde{\delta^2}.
\label{eqn:delta21b}
\end{equation}
We motivated this form by keeping all bias terms needed for the power spectrum at 1-loop order and by, then, arguing on physical grounds (and with simulations of reionization) that certain terms are small over much of the observationally relevant wavenumber range.  We further showed that linear theory is likely sufficient for computing $\delta^2$.  Each bias coefficient in this model has physical meanings, namely
 \begin{itemize}
 \item $b_1 \approx \bar x_{\rm H} - (1-\bar x_{\rm H}) b_{S,1}$ for much of reionization, where $\bar x_{\rm H}$ is the global average of the neutral fraction and $b_{S, 1}$ is the linear bias of the sources.  Thus, $b_1$ constrains the sources and global neutral fraction.  This formula qualitatively fails when recombinations alter the morphology of reionization (as models suggest happens towards the end; \cite{furlanetto05}), decreasing $b_1$.  In our reference simulations, a breakdown of this expression for $b_1$ occurs near the end of reionization ($\bar{x}_H \lesssim 0.2$ in two of our three simulations, and $\bar{x}_H \lesssim 0.5$ in the Abundant Sinks simulation).
 \item $b_2$ is often driven by the coupling of fluctuations in the ionization field to those in the source radiation field.  This coefficient is often large because reionization is patchy, with its inclusion having a dramatic affect on the accuracy of our expansion.  Additionally, all of our simulations show an extended epoch where $b_1\approx 0$ such that $b_2 \delta^2$ is the dominant term (which occurs when $\bar x_{\rm H} \approx [1-\bar x_{\rm H}] b_{S,1}$).  
 \item $R_{\rm eff}$ encodes the effective ionized bubble size and is only applicable on perturbative scales towards the latter half of reionization, being responsible for the well known flattening of the signal (and this flattening indicates perturbation theory is failing as should happen when $k R_{\rm eff} \sim 1$).  Its value in our simulations is considerably smaller than the characteristic bubble size in semi-analytic reionization models.
 \end{itemize}
   We found that $b_1$ is typically constrained by a measurement of the 21cm power spectrum with modest signal-to-noise ratios in just a few band-power bins.  However, the requirements are more stringent to measure the other two parameters.   Our perturbative expansion is agnostic about the sources and sinks of ionizing photons during reionization, with this ultraviolet physics encapsulated in the values of these coefficients.
   
  The accuracy of this expansion varies significantly when tested against our three reionization simulations and, even when considering a particular simulation, as a function of time, often describing reliably $k\lesssim 0.2-0.5~h\,$Mpc$^{-1}$ -- much of the wavenumber range potentially probed by upcoming radio telescopes.  The error power spectrum is the largest and, hence, it is the least successful in our High Mass simulation.   The model's error power spectrum for all three simulations is generally neither white nor does it scale as strongly with wavenumber as omitted higher-order terms, suggesting that the errors owe to non-perturbative effects, likely arising from the largest bubbles.   We also note that the Fiducial and High Mass simulations almost certainly do not resolve the dense clumps that are predicted to limit the maximum bubble sizes during reionization \citep{furlanetto05, 2014MNRAS.440.1662S}.  
   Since the largest bubbles lead to the largest non-perturbative errors, the actual 21cm signal may be more perturbative than the simulated signals.  
   
 We also investigated more general bias expansions.  First, we fit each term having a different time dependence with a separate coefficient such that, e.g., the linear matter overdensity $\delta^{(1)}$ has a different coefficient than the second-order matter overdensity $\delta^{(2)}$ (plus some accounting to evaluate these fields along Lagrangian paths).  We found that this separation only moderately improves the goodness of fits at the mode level, fortunately, as such an expansion has too much freedom to be useful.  Our simpler {\it Minimal Model} is a good approximation because the matter density is relatively linear on scales where our theory applies -- the dominant nonlinearity in the 21cm signal owes to the radiative transfer from highly biased sources.  (Shot noise from the numerical computation of these higher-order-in-$\delta^{(1)}$ terms also limits the  effectiveness of our method at reducing the residuals with these terms.)  Second, we fit the simulations with $k$-dependent transfer functions multiplying $[\delta]$ and $[\delta^2]$, motivated by the largeness of the cross correlation coefficient between $\delta_{21}$ and $\delta^{(1)}$ at higher wavenumbers than where our {\it Minimal Model} is successful.  We found that this additional freedom improved the fits, but only at wavenumbers where the model errors were already substantial such that a fully nonlinear model is likely needed.


There are a several avenues for future work.  This model can be used to study other large-scale statistics of the 21cm field (in addition to the power spectrum) as well as other large-scale reionization observables.  
  In addition, we have worked in the limit where spin temperature ($T_S$) fluctuations contribute negligibly to the 21cm signal.  A different theoretical approach is required for the contribution of $T_S$ fluctuations because the mean free path of the X-ray and ultraviolet photons, which respectively heat the gas and couple its temperature to $T_S$, tend to be larger than the scales potentially probed by 21cm observations.  {\refereeedit (We suspect spin temperature fluctuations can be incorporated by expanding in inverse powers of $k$, resulting in first the additional term $(\lambda_{\rm mfp} k)^{-1} \delta^{(1)}$, where $\lambda_{\rm mfp}$ is the effective mean free path of the photons that affect $T_S$.)}  Additionally, our model has not included redshift-space distortions.  While redshift-space distortions would likely complicate somewhat the expansion, we expect that these distortions would not significantly impair the bias expansion's accuracy because the radiative transfer is the largest nonlinearity.  Finally, and most importantly, additional reionization simulations (and/or semi-analytic realizations of reionization) to understand the proposed bias model and where it applies are essential for using this model as an interpretive tool.

\paragraph{Acknowledgments}  We thank Adam Lidz for storing the reionization simulations on his computers for so many years and Jonnie Pober for the HERA sensitivity forecasts.  We also thank Marcelo Alvarez and Vid Ir\v si\v c for useful conversations, and  Vid Ir\v si\v c and Adam Lidz for helpful comments on a draft version.  This work is supported by NSF award AST1614439, NASA award NNX17AH68G, NASA/HST award HST-AR-15013.005-A, and by the Alfred P. Sloan foundation.


%

\bibliography{References}

\providecommand{\href}[2]{#2}\begingroup\raggedright\begin{thebibliography}{10}

\bibitem{mcquinn-review}
M.~{McQuinn}, {\it {The Evolution of the Intergalactic Medium}},  {\em \araa}
  {\bf 54} (Sept., 2016) 313--362,
  [\href{http://xxx.lanl.gov/abs/1512.0008}{{\tt arXiv:1512.0008}}].

\bibitem{mcgreer15}
I.~D. {McGreer}, A.~{Mesinger}, and V.~{D'Odorico}, {\it {Model-independent
  evidence in favour of an end to reionization by $z {\approx} 6$}},  {\em
  \mnras} {\bf 447} (Feb., 2015) 499--505,
  [\href{http://xxx.lanl.gov/abs/1411.5375}{{\tt arXiv:1411.5375}}].

\bibitem{2016arXiv160503507P}
{Planck Collaboration}, R.~{Adam}, N.~{Aghanim}, M.~{Ashdown}, J.~{Aumont},
  C.~{Baccigalupi}, M.~{Ballardini}, A.~J. {Banday}, R.~B. {Barreiro}, and
  N.~e.~a. {Bartolo}, {\it {Planck intermediate results. XLVII. Planck
  constraints on reionization history}},  {\em ArXiv e-prints} (May, 2016)
  [\href{http://xxx.lanl.gov/abs/1605.0350}{{\tt arXiv:1605.0350}}].

\bibitem{furl-models}
S.~R. {Furlanetto}, M.~{McQuinn}, and L.~{Hernquist}, {\it {Characteristic
  scales during reionization}},  {\em \mnras} {\bf 365} (Jan., 2006) 115--126,
  [\href{http://xxx.lanl.gov/abs/astro-ph/0507524}{{\tt astro-ph/0507524}}].

\bibitem{mcquinn07}
M.~{McQuinn}, A.~{Lidz}, O.~{Zahn}, S.~{Dutta}, L.~{Hernquist}, and
  M.~{Zaldarriaga}, {\it {The morphology of HII regions during reionization}},
  {\em \mnras} {\bf 377} (May, 2007) 1043--1063,
  [\href{http://xxx.lanl.gov/abs/astro-ph/0610094}{{\tt astro-ph/0610094}}].

\bibitem{2017MNRAS.469.4283K}
G.~{Kulkarni}, T.~R. {Choudhury}, E.~{Puchwein}, and M.~G. {Haehnelt}, {\it
  {Large 21-cm signals from AGN-dominated reionization}},  {\em \mnras} {\bf
  469} (Aug., 2017) 4283--4291, [\href{http://xxx.lanl.gov/abs/1701.0440}{{\tt
  arXiv:1701.0440}}].

\bibitem{ciardi06}
B.~{Ciardi}, E.~{Scannapieco}, F.~{Stoehr}, A.~{Ferrara}, I.~T. {Iliev}, and
  P.~R. {Shapiro}, {\it {The effect of minihaloes on cosmic reionization}},
  {\em \mnras} {\bf 366} (Feb., 2006) 689--696,
  [\href{http://xxx.lanl.gov/abs/astro-ph/0511623}{{\tt astro-ph/0511623}}].

\bibitem{2009MNRAS.394..960C}
T.~R. {Choudhury}, M.~G. {Haehnelt}, and J.~{Regan}, {\it {Inside-out or
  outside-in: the topology of reionization in the photon-starved regime
  suggested by Ly{$\alpha$} forest data}},  {\em \mnras} {\bf 394} (Apr., 2009)
  960--977, [\href{http://xxx.lanl.gov/abs/0806.1524}{{\tt arXiv:0806.1524}}].

\bibitem{furlanetto07b}
S.~R. {Furlanetto} and S.~P. {Oh}, {\it {Inhomogeneous Helium Reionization and
  the Equation of State of the Intergalactic Medium}},  {\em \apj} {\bf 682}
  (July, 2008) 14--28, [\href{http://xxx.lanl.gov/abs/0711.0751}{{\tt
  arXiv:0711.0751}}].

\bibitem{barkana04}
R.~{Barkana} and A.~{Loeb}, {\it {A Method for Separating the Physics from the
  Astrophysics of High-Redshift 21 Centimeter Fluctuations}},  {\em \apjl} {\bf
  624} (May, 2005) L65--L68,
  [\href{http://xxx.lanl.gov/abs/astro-ph/0409572}{{\tt astro-ph/0409572}}].

\bibitem{mcquinn06}
M.~{McQuinn}, O.~{Zahn}, M.~{Zaldarriaga}, L.~{Hernquist}, and S.~R.
  {Furlanetto}, {\it {Cosmological Parameter Estimation Using 21 cm Radiation
  from the Epoch of Reionization}},  {\em \apj} {\bf 653} (Dec., 2006)
  815--834, [\href{http://xxx.lanl.gov/abs/astro-ph/0512263}{{\tt
  astro-ph/0512263}}].

\bibitem{2008PhRvD..78b3529M}
Y.~{Mao}, M.~{Tegmark}, M.~{McQuinn}, M.~{Zaldarriaga}, and O.~{Zahn}, {\it
  {How accurately can 21cm tomography constrain cosmology?}},  {\em \prd} {\bf
  78} (July, 2008) 023529, [\href{http://xxx.lanl.gov/abs/0802.1710}{{\tt
  arXiv:0802.1710}}].

\bibitem{2013MNRAS.433.2900D}
A.~{D'Aloisio}, J.~{Zhang}, P.~R. {Shapiro}, and Y.~{Mao}, {\it {The
  scale-dependent signature of primordial non-Gaussianity in the large-scale
  structure of cosmic reionization}},  {\em \mnras} {\bf 433} (Aug., 2013)
  2900--2919, [\href{http://xxx.lanl.gov/abs/1304.6411}{{\tt
  arXiv:1304.6411}}].

\bibitem{2013PhRvD..88h1303M}
Y.~{Mao}, A.~{D'Aloisio}, J.~{Zhang}, and P.~R. {Shapiro}, {\it {Primordial
  non-Gaussianity estimation using 21 cm tomography from the epoch of
  reionization}},  {\em \prd} {\bf 88} (Oct., 2013) 081303,
  [\href{http://xxx.lanl.gov/abs/1305.0313}{{\tt arXiv:1305.0313}}].

\bibitem{2013PhRvD..88b3534L}
A.~{Lidz}, E.~J. {Baxter}, P.~{Adshead}, and S.~{Dodelson}, {\it {Primordial
  non-Gaussianity and reionization}},  {\em \prd} {\bf 88} (July, 2013) 023534,
  [\href{http://xxx.lanl.gov/abs/1304.8049}{{\tt arXiv:1304.8049}}].

\bibitem{madau97}
P.~{Madau}, A.~{Meiksin}, and M.~J. {Rees}, {\it {21 Centimeter Tomography of
  the Intergalactic Medium at High Redshift}},  {\em \apj} {\bf 475} (Feb.,
  1997) 429--444, [\href{http://xxx.lanl.gov/abs/astro-ph/9608010}{{\tt
  astro-ph/9608010}}].

\bibitem{ciardi03-21cm}
B.~{Ciardi} and P.~{Madau}, {\it {Probing beyond the Epoch of Hydrogen
  Reionization with 21 Centimeter Radiation}},  {\em \apj} {\bf 596} (Oct.,
  2003) 1--8, [\href{http://xxx.lanl.gov/abs/astro-ph/0303249}{{\tt
  astro-ph/0303249}}].

\bibitem{zaldarriaga04}
M.~{Zaldarriaga}, S.~R. {Furlanetto}, and L.~{Hernquist}, {\it {21 Centimeter
  Fluctuations from Cosmic Gas at High Redshifts}},  {\em \apj} {\bf 608}
  (June, 2004) 622--635, [\href{http://xxx.lanl.gov/abs/astro-ph/0311514}{{\tt
  astro-ph/0311514}}].

\bibitem{furl-rev}
S.~R. {Furlanetto}, S.~P. {Oh}, and F.~H. {Briggs}, {\it {Cosmology at low
  frequencies: The 21 cm transition and the high-redshift Universe}},  {\em
  Physics Reports} {\bf 433} (Oct., 2006) 181--301,
  [\href{http://xxx.lanl.gov/abs/astro-ph/0608032}{{\tt astro-ph/0608032}}].

\bibitem{morales03}
M.~F. {Morales} and J.~{Hewitt}, {\it {Toward Epoch of Reionization
  Measurements with Wide-Field Radio Observations}},  {\em \apj} {\bf 615}
  (Nov., 2004) 7--18, [\href{http://xxx.lanl.gov/abs/astro-ph/0312437}{{\tt
  astro-ph/0312437}}].

\bibitem{bowman05}
J.~D. {Bowman}, M.~F. {Morales}, and J.~N. {Hewitt}, {\it {The Sensitivity of
  First-Generation Epoch of Reionization Observatories and Their Potential for
  Differentiating Theoretical Power Spectra}},  {\em \apj} {\bf 638} (Feb.,
  2006) 20--26, [\href{http://xxx.lanl.gov/abs/astro-ph/0507357}{{\tt
  astro-ph/0507357}}].

\bibitem{2011PhRvD..83j3006L}
A.~{Liu} and M.~{Tegmark}, {\it {A method for 21 cm power spectrum estimation
  in the presence of foregrounds}},  {\em \prd} {\bf 83} (May, 2011) 103006,
  [\href{http://xxx.lanl.gov/abs/1103.0281}{{\tt arXiv:1103.0281}}].

\bibitem{parsons12}
A.~{Parsons}, J.~{Pober}, M.~{McQuinn}, D.~{Jacobs}, and J.~{Aguirre}, {\it {A
  Sensitivity and Array-configuration Study for Measuring the Power Spectrum of
  21 cm Emission from Reionization}},  {\em \apj} {\bf 753} (July, 2012) 81,
  [\href{http://xxx.lanl.gov/abs/1103.2135}{{\tt arXiv:1103.2135}}].

\bibitem{dillon15}
J.~S. {Dillon} et~al., {\it {Empirical covariance modeling for 21 cm power
  spectrum estimation: A method demonstration and new limits from early
  Murchison Widefield Array 128-tile data}},  {\em \prd} {\bf 91} (June, 2015)
  123011, [\href{http://xxx.lanl.gov/abs/1506.0102}{{\tt arXiv:1506.0102}}].

\bibitem{2016ApJ...819....8P}
J.~C. {Pober} et~al., {\it {The Importance of Wide-field Foreground Removal for
  21 cm Cosmology: A Demonstration with Early MWA Epoch of Reionization
  Observations}},  {\em \apj} {\bf 819} (Mar., 2016) 8,
  [\href{http://xxx.lanl.gov/abs/1601.0617}{{\tt arXiv:1601.0617}}].

\bibitem{2016MNRAS.461.3135B}
N.~{Barry}, B.~{Hazelton}, I.~{Sullivan}, M.~F. {Morales}, and J.~C. {Pober},
  {\it {Calibration requirements for detecting the 21 cm epoch of reionization
  power spectrum and implications for the SKA}},  {\em \mnras} {\bf 461}
  (Sept., 2016) 3135--3144, [\href{http://xxx.lanl.gov/abs/1603.0060}{{\tt
  arXiv:1603.0060}}].

\bibitem{2017MNRAS.470.1849E}
A.~{Ewall-Wice}, J.~S. {Dillon}, A.~{Liu}, and J.~{Hewitt}, {\it {The impact of
  modelling errors on interferometer calibration for 21 cm power spectra}},
  {\em \mnras} {\bf 470} (Sept., 2017) 1849--1870,
  [\href{http://xxx.lanl.gov/abs/1610.0268}{{\tt arXiv:1610.0268}}].

\bibitem{2016ApJ...833..102B}
A.~P. {Beardsley} et~al., {\it {First Season MWA EoR Power spectrum Results at
  Redshift 7}},  {\em \apj} {\bf 833} (Dec., 2016) 102,
  [\href{http://xxx.lanl.gov/abs/1608.0628}{{\tt arXiv:1608.0628}}].

\bibitem{2017ApJ...838...65P}
A.~H. {Patil}, S.~{Yatawatta}, L.~V.~E. {Koopmans}, A.~G. {de Bruyn}, M.~A.
  {Brentjens}, S.~{Zaroubi}, K.~M.~B. {Asad}, M.~{Hatef}, and et~al., {\it
  {Upper Limits on the 21 cm Epoch of Reionization Power Spectrum from One
  Night with LOFAR}},  {\em \apj} {\bf 838} (Mar., 2017) 65,
  [\href{http://xxx.lanl.gov/abs/1702.0867}{{\tt arXiv:1702.0867}}].

\bibitem{2017PASP..129d5001D}
D.~R. {DeBoer} et~al., {\it {Hydrogen Epoch of Reionization Array (HERA)}},
  {\em \pasp} {\bf 129} (Apr., 2017) 045001,
  [\href{http://xxx.lanl.gov/abs/1606.0747}{{\tt arXiv:1606.0747}}].

\bibitem{2015aska.confE...1K}
L.~{Koopmans}, J.~{Pritchard}, G.~{Mellema}, J.~{Aguirre}, K.~{Ahn},
  R.~{Barkana}, I.~{van Bemmel}, G.~{Bernardi}, and et~al., {\it {The Cosmic
  Dawn and Epoch of Reionisation with SKA}},  {\em Advancing Astrophysics with
  the Square Kilometre Array (AASKA14)} (Apr., 2015) 1,
  [\href{http://xxx.lanl.gov/abs/1505.0756}{{\tt arXiv:1505.0756}}].

\bibitem{gnedin97}
N.~Y. {Gnedin} and J.~P. {Ostriker}, {\it {Reionization of the Universe and the
  Early Production of Metals}},  {\em \apj} {\bf 486} (Sept., 1997) 581--598,
  [\href{http://xxx.lanl.gov/abs/astro-ph/9612127}{{\tt astro-ph/9612127}}].

\bibitem{ciardi03-sim}
B.~{Ciardi}, F.~{Stoehr}, and S.~D.~M. {White}, {\it {Simulating intergalactic
  medium reionization}},  {\em \mnras} {\bf 343} (Aug., 2003) 1101--1109,
  [\href{http://xxx.lanl.gov/abs/astro-ph/0301293}{{\tt astro-ph/0301293}}].

\bibitem{iliev05}
I.~T. {Iliev}, G.~{Mellema}, U.-L. {Pen}, H.~{Merz}, P.~R. {Shapiro}, and M.~A.
  {Alvarez}, {\it {Simulating cosmic reionization at large scales - I. The
  geometry of reionization}},  {\em \mnras} {\bf 369} (July, 2006) 1625--1638,
  [\href{http://xxx.lanl.gov/abs/astro-ph/0512187}{{\tt astro-ph/0512187}}].

\bibitem{trac07}
H.~{Trac} and R.~{Cen}, {\it {Radiative Transfer Simulations of Cosmic
  Reionization. I. Methodology and Initial Results}},  {\em \apj} {\bf 671}
  (Dec., 2007) 1--13, [\href{http://xxx.lanl.gov/abs/astro-ph/0612406}{{\tt
  astro-ph/0612406}}].

\bibitem{2014ApJ...793...30G}
N.~Y. {Gnedin} and A.~A. {Kaurov}, {\it {Cosmic Reionization on Computers. II.
  Reionization History and Its Back-reaction on Early Galaxies}},  {\em \apj}
  {\bf 793} (Sept., 2014) 30, [\href{http://xxx.lanl.gov/abs/1403.4251}{{\tt
  arXiv:1403.4251}}].

\bibitem{2017MNRAS.466..960P}
A.~H. {Pawlik}, A.~{Rahmati}, J.~{Schaye}, M.~{Jeon}, and C.~{Dalla Vecchia},
  {\it {The Aurora radiation-hydrodynamical simulations of reionization:
  calibration and first results}},  {\em \mnras} {\bf 466} (Apr., 2017)
  960--973, [\href{http://xxx.lanl.gov/abs/1603.0003}{{\tt arXiv:1603.0003}}].

\bibitem{miralda00}
J.~{Miralda-Escud{\'e}}, M.~{Haehnelt}, and M.~J. {Rees}, {\it {Reionization of
  the Inhomogeneous Universe}},  {\em \apj} {\bf 530} (Feb., 2000) 1--16,
  [\href{http://xxx.lanl.gov/abs/astro-ph/9812306}{{\tt astro-ph/9812306}}].

\bibitem{gnedin00}
N.~Y. {Gnedin}, {\it {Effect of Reionization on Structure Formation in the
  Universe}},  {\em \apj} {\bf 542} (Oct., 2000) 535--541,
  [\href{http://xxx.lanl.gov/abs/astro-ph/0002151}{{\tt astro-ph/0002151}}].

\bibitem{shapiro03}
P.~R. {Shapiro}, I.~T. {Iliev}, and A.~C. {Raga}, {\it {Photoevaporation of
  cosmological minihaloes during reionization}},  {\em \mnras} {\bf 348} (Mar.,
  2004) 753--782, [\href{http://xxx.lanl.gov/abs/astro-ph/0307266}{{\tt
  astro-ph/0307266}}].

\bibitem{mesinger11}
A.~{Mesinger}, S.~{Furlanetto}, and R.~{Cen}, {\it {21CMFAST: a fast,
  seminumerical simulation of the high-redshift 21-cm signal}},  {\em \mnras}
  {\bf 411} (Feb., 2011) 955--972,
  [\href{http://xxx.lanl.gov/abs/1003.3878}{{\tt arXiv:1003.3878}}].

\bibitem{2010MNRAS.406.2421S}
M.~G. {Santos}, L.~{Ferramacho}, M.~B. {Silva}, A.~{Amblard}, and A.~{Cooray},
  {\it {Fast large volume simulations of the 21-cm signal from the reionization
  and pre-reionization epochs}},  {\em \mnras} {\bf 406} (Aug., 2010)
  2421--2432, [\href{http://xxx.lanl.gov/abs/0911.2219}{{\tt
  arXiv:0911.2219}}].

\bibitem{furlanetto05}
S.~R. {Furlanetto} and S.~P. {Oh}, {\it {Taxing the rich: recombinations and
  bubble growth during reionization}},  {\em \mnras} {\bf 363} (Nov., 2005)
  1031--1048, [\href{http://xxx.lanl.gov/abs/astro-ph/0505065}{{\tt
  astro-ph/0505065}}].

\bibitem{alvarez12}
M.~A. {Alvarez} and T.~{Abel}, {\it {The Effect of Absorption Systems on Cosmic
  Reionization}},  {\em \apj} {\bf 747} (Mar., 2012) 126,
  [\href{http://xxx.lanl.gov/abs/1003.6132}{{\tt arXiv:1003.6132}}].

\bibitem{sobacchi14}
E.~{Sobacchi} and A.~{Mesinger}, {\it {Inhomogeneous recombinations during
  cosmic reionization}},  {\em \mnras} {\bf 440} (May, 2014) 1662--1673,
  [\href{http://xxx.lanl.gov/abs/1402.2298}{{\tt arXiv:1402.2298}}].

\bibitem{zahn06}
O.~{Zahn}, A.~{Lidz}, M.~{McQuinn}, S.~{Dutta}, L.~{Hernquist},
  M.~{Zaldarriaga}, and S.~R. {Furlanetto}, {\it {Simulations and Analytic
  Calculations of Bubble Growth during Hydrogen Reionization}},  {\em \apj}
  {\bf 654} (Jan., 2007) 12--26,
  [\href{http://xxx.lanl.gov/abs/astro-ph/0604177}{{\tt astro-ph/0604177}}].

\bibitem{zahn11}
O.~{Zahn}, A.~{Mesinger}, M.~{McQuinn}, H.~{Trac}, R.~{Cen}, and L.~E.
  {Hernquist}, {\it {Comparison of reionization models: radiative transfer
  simulations and approximate, seminumeric models}},  {\em \mnras} {\bf 414}
  (June, 2011) 727--738, [\href{http://xxx.lanl.gov/abs/1003.3455}{{\tt
  arXiv:1003.3455}}].

\bibitem{2014MNRAS.443.2843M}
S.~{Majumdar}, G.~{Mellema}, K.~K. {Datta}, H.~{Jensen}, T.~R. {Choudhury},
  S.~{Bharadwaj}, and M.~M. {Friedrich}, {\it {On the use of seminumerical
  simulations in predicting the 21-cm signal from the epoch of reionization}},
  {\em \mnras} {\bf 443} (Oct., 2014) 2843--2861,
  [\href{http://xxx.lanl.gov/abs/1403.0941}{{\tt arXiv:1403.0941}}].

\bibitem{2007MNRAS.375..324Z}
J.~{Zhang}, L.~{Hui}, and Z.~{Haiman}, {\it {A linear perturbation theory of
  inhomogeneous reionization}},  {\em \mnras} {\bf 375} (Feb., 2007) 324--336,
  [\href{http://xxx.lanl.gov/abs/astro-ph/0607628}{{\tt astro-ph/0607628}}].

\bibitem{2015PhRvD..91h3015M}
Y.~{Mao}, A.~{D'Aloisio}, B.~D. {Wandelt}, J.~{Zhang}, and P.~R. {Shapiro},
  {\it {Linear perturbation theory of reionization in position space:
  Cosmological radiative transfer along the light cone}},  {\em \prd} {\bf 91}
  (Apr., 2015) 083015, [\href{http://xxx.lanl.gov/abs/1411.7022}{{\tt
  arXiv:1411.7022}}].

\bibitem{2007ApJ...659..865L}
A.~{Lidz}, O.~{Zahn}, M.~{McQuinn}, M.~{Zaldarriaga}, S.~{Dutta}, and
  L.~{Hernquist}, {\it {Higher Order Contributions to the 21 cm Power
  Spectrum}},  {\em \apj} {\bf 659} (Apr., 2007) 865--876,
  [\href{http://xxx.lanl.gov/abs/astro-ph/0610054}{{\tt astro-ph/0610054}}].

\bibitem{mcquinnLya}
M.~{McQuinn}, L.~{Hernquist}, M.~{Zaldarriaga}, and S.~{Dutta}, {\it {Studying
  reionization with Ly{$\alpha$} emitters}},  {\em \mnras} {\bf 381} (Oct.,
  2007) 75--96, [\href{http://xxx.lanl.gov/abs/0704.2239}{{\tt
  arXiv:0704.2239}}].

\bibitem{2018arXiv180202578H}
K.~{Hoffmann}, Y.~{Mao}, H.~{Mo}, and B.~D. {Wandelt}, {\it {Signatures of
  Cosmic Reionization on the 21cm 2- and 3-point Correlation Function I:
  Quadratic Bias Modeling}},  {\em ArXiv e-prints} (Feb., 2018)
  [\href{http://xxx.lanl.gov/abs/1802.0257}{{\tt arXiv:1802.0257}}].

\bibitem{lidz08}
A.~{Lidz}, O.~{Zahn}, M.~{McQuinn}, M.~{Zaldarriaga}, and L.~{Hernquist}, {\it
  {Detecting the Rise and Fall of 21 cm Fluctuations with the Murchison
  Widefield Array}},  {\em \apj} {\bf 680} (June, 2008) 962--974,
  [\href{http://xxx.lanl.gov/abs/0711.4373}{{\tt arXiv:0711.4373}}].

\bibitem{2017MNRAS.472.1576F}
S.~R. {Furlanetto}, J.~{Mirocha}, R.~H. {Mebane}, and G.~{Sun}, {\it {A
  minimalist feedback-regulated model for galaxy formation during the epoch of
  reionization}},  {\em \mnras} {\bf 472} (Dec., 2017) 1576--1592,
  [\href{http://xxx.lanl.gov/abs/1611.0116}{{\tt arXiv:1611.0116}}].

\bibitem{becker15}
G.~D. {Becker}, J.~S. {Bolton}, P.~{Madau}, M.~{Pettini}, E.~V. {Ryan-Weber},
  and B.~P. {Venemans}, {\it {Evidence of patchy hydrogen reionization from an
  extreme Ly{$\alpha$} trough below redshift six}},  {\em \mnras} {\bf 447}
  (Mar., 2015) 3402--3419, [\href{http://xxx.lanl.gov/abs/1407.4850}{{\tt
  arXiv:1407.4850}}].

\bibitem{2016MNRAS.460.1328D}
F.~B. {Davies} and S.~R. {Furlanetto}, {\it {Large fluctuations in the
  hydrogen-ionizing background and mean free path following the epoch of
  reionization}},  {\em \mnras} {\bf 460} (Aug., 2016) 1328--1339,
  [\href{http://xxx.lanl.gov/abs/1509.0713}{{\tt arXiv:1509.0713}}].

\bibitem{daloisio-gal}
A.~{D'Aloisio}, M.~{McQuinn}, F.~B. {Davies}, and S.~R. {Furlanetto}, {\it
  {Large fluctuations in the high-redshift metagalactic ionizing background}},
  {\em \mnras} {\bf 473} (Jan., 2018) 560--575,
  [\href{http://xxx.lanl.gov/abs/1611.0271}{{\tt arXiv:1611.0271}}].

\bibitem{2014MNRAS.440.1662S}
E.~{Sobacchi} and A.~{Mesinger}, {\it {Inhomogeneous recombinations during
  cosmic reionization}},  {\em \mnras} {\bf 440} (May, 2014) 1662--1673,
  [\href{http://xxx.lanl.gov/abs/1402.2298}{{\tt arXiv:1402.2298}}].

\bibitem{mirocha17}
J.~{Mirocha}, S.~R. {Furlanetto}, and G.~{Sun}, {\it {The global 21-cm signal
  in the context of the high- z galaxy luminosity function}},  {\em \mnras}
  {\bf 464} (Jan., 2017) 1365--1379,
  [\href{http://xxx.lanl.gov/abs/1607.0038}{{\tt arXiv:1607.0038}}].

\bibitem{2016arXiv161109787D}
V.~{Desjacques}, D.~{Jeong}, and F.~{Schmidt}, {\it {Large-Scale Galaxy Bias}},
   {\em ArXiv e-prints} (Nov., 2016)
  [\href{http://xxx.lanl.gov/abs/1611.0978}{{\tt arXiv:1611.0978}}].

\bibitem{2009JCAP...08..020M}
P.~{McDonald} and A.~{Roy}, {\it {Clustering of dark matter tracers:
  generalizing bias for the coming era of precision LSS}},  {\em \jcap} {\bf 8}
  (Aug., 2009) 020, [\href{http://xxx.lanl.gov/abs/0902.0991}{{\tt
  arXiv:0902.0991}}].

\bibitem{2014JCAP...08..056A}
V.~{Assassi}, D.~{Baumann}, D.~{Green}, and M.~{Zaldarriaga}, {\it
  {Renormalized halo bias}},  {\em \jcap} {\bf 8} (Aug., 2014) 056,
  [\href{http://xxx.lanl.gov/abs/1402.5916}{{\tt arXiv:1402.5916}}].

\bibitem{2015JCAP...11..007S}
L.~{Senatore}, {\it {Bias in the effective field theory of large scale
  structures}},  {\em \jcap} {\bf 11} (Nov., 2015) 007,
  [\href{http://xxx.lanl.gov/abs/1406.7843}{{\tt arXiv:1406.7843}}].

\bibitem{mcdonald06}
P.~{McDonald}, {\it {Clustering of dark matter tracers: Renormalizing the bias
  parameters}},  {\em \prd} {\bf 74} (Nov., 2006) 103512,
  [\href{http://xxx.lanl.gov/abs/astro-ph/0609413}{{\tt astro-ph/0609413}}].

\bibitem{daloisio18}
A.~{D'Aloisio}, M.~{McQuinn}, O.~{Maupin}, F.~B. {Davies}, H.~{Trac},
  S.~{Fuller}, and P.~R. {Upton Sanderbeck}, {\it {Heating of the Intergalactic
  Medium by Hydrogen Reionization}},  {\em ArXiv e-prints} (July, 2018)
  [\href{http://xxx.lanl.gov/abs/1807.0928}{{\tt arXiv:1807.0928}}].

\bibitem{2016JCAP...03..007B}
T.~{Baldauf}, E.~{Schaan}, and M.~{Zaldarriaga}, {\it {On the reach of
  perturbative methods for dark matter density fields}},  {\em \jcap} {\bf 3}
  (Mar., 2016) 007, [\href{http://xxx.lanl.gov/abs/1507.0225}{{\tt
  arXiv:1507.0225}}].

\bibitem{carrasco12}
J.~J.~M. {Carrasco}, M.~P. {Hertzberg}, and L.~{Senatore}, {\it {The effective
  field theory of cosmological large scale structures}},  {\em Journal of High
  Energy Physics} {\bf 9} (Sept., 2012) 82,
  [\href{http://xxx.lanl.gov/abs/1206.2926}{{\tt arXiv:1206.2926}}].

\bibitem{mcquinnwhite}
M.~{McQuinn} and M.~{White}, {\it {Cosmological perturbation theory in 1+1
  dimensions}},  {\em \jcap} {\bf 1} (Jan., 2016) 043,
  [\href{http://xxx.lanl.gov/abs/1502.0738}{{\tt arXiv:1502.0738}}].

\bibitem{2014ApJ...782...66P}
J.~C. {Pober}, A.~{Liu}, J.~S. {Dillon}, J.~E. {Aguirre}, J.~D. {Bowman}, R.~F.
  {Bradley}, C.~L. {Carilli}, D.~R. {DeBoer}, J.~N. {Hewitt}, D.~C. {Jacobs},
  M.~{McQuinn}, M.~F. {Morales}, A.~R. {Parsons}, M.~{Tegmark}, and D.~J.
  {Werthimer}, {\it {What Next-generation 21 cm Power Spectrum Measurements can
  Teach us About the Epoch of Reionization}},  {\em \apj} {\bf 782} (Feb.,
  2014) 66, [\href{http://xxx.lanl.gov/abs/1310.7031}{{\tt arXiv:1310.7031}}].

\bibitem{2001MNRAS.323....1S}
R.~K. {Sheth}, H.~J. {Mo}, and G.~{Tormen}, {\it {Ellipsoidal collapse and an
  improved model for the number and spatial distribution of dark matter
  haloes}},  {\em \mnras} {\bf 323} (May, 2001) 1--12,
  [\href{http://xxx.lanl.gov/abs/astro-ph/9907024}{{\tt astro-ph/9907024}}].

\bibitem{cooray02}
A.~{Cooray} and R.~{Sheth}, {\it {Halo models of large scale structure}},  {\em
  \physrep} {\bf 372} (Dec., 2002) 1--129,
  [\href{http://xxx.lanl.gov/abs/astro-ph/0206508}{{\tt astro-ph/0206508}}].

\bibitem{mcquinn11}
M.~{McQuinn}, L.~{Hernquist}, A.~{Lidz}, and M.~{Zaldarriaga}, {\it {The
  signatures of large-scale temperature and intensity fluctuations in the Lyman
  {$\alpha$} forest}},  {\em \mnras} {\bf 415} (July, 2011) 977--992,
  [\href{http://xxx.lanl.gov/abs/1010.5250}{{\tt arXiv:1010.5250}}].

\bibitem{sheth99}
R.~K. Sheth and G.~Lemson, {\it The forest of merger history trees associated
  with the formation of dark matter halos},
  \href{http://xxx.lanl.gov/abs/astro-ph/9805322}{{\tt astro-ph/9805322}}.

\bibitem{furlanetto04a}
S.~R. {Furlanetto}, M.~{Zaldarriaga}, and L.~{Hernquist}, {\it {The Growth of H
  II Regions During Reionization}},  {\em \apj} {\bf 613} (Sept., 2004) 1--15.

\bibitem{tassev13}
S.~{Tassev} and M.~{Zaldarriaga}, {\it {The mildly non-linear regime of
  structure formation}},  {\em \jcap} {\bf 4} (Apr., 2012) 13,
  [\href{http://xxx.lanl.gov/abs/1109.4939}{{\tt arXiv:1109.4939}}].

\end{thebibliography}\endgroup
 
\newpage

\end{document}